\documentclass[journal]{IEEEtran}
\usepackage{cite}
\usepackage{amsmath,amssymb,amsfonts,amsthm}
\usepackage{bm}
\usepackage{graphicx}
\usepackage{url}
\usepackage[hidelinks]{hyperref}
\usepackage{algorithm}
\usepackage{algorithmic}
\usepackage[caption=false,font=footnotesize]{subfig}
\usepackage{xcolor}
\usepackage{enumitem}
\newtheorem{lemma}{Lemma}
\newtheorem{theorem}{Theorem}

\allowdisplaybreaks

\title{Statistical Ambiguity-Function-Aware\\Joint Beamforming and Subcarrier Allocation for\\MIMO-OFDM ISAC under Random Symbols}

\author{
Wonill Kim,~\IEEEmembership{Student Member,~IEEE,}
Yumin Kim,~\IEEEmembership{Student Member,~IEEE,}
Hyeonsu Lyu,~\IEEEmembership{Member,~IEEE,}
\newline and Hyun Jong Yang,~\IEEEmembership{Senior Member,~IEEE}\thanks{
Wonill Kim, Yumin Kim, and Hyun Jong Yang are with the Department of Electrical and Computer Engineering, Seoul National University, Korea (email: \{kwi030104, yumin0107, hjyang\}@snu.ac.kr).
Hyeonsu Lyu and Hyun Jong Yang are with the Institute of New Media and Communications, Seoul National University, Korea (email: hs.lyu@snu.ac.kr).
}
}

\begin{document}

\maketitle

\begin{abstract}
Ambiguity function (AF) characterizes the delay–Doppler sidelobe structure that governs target resolution and masking in radar sensing.
In integrated sensing and communications (ISAC), however, random communication symbols make the AF data-dependent, complicating its characterization and incorporation into transmit design.
This raises a fundamental question: which AF characteristic can be effectively controlled through beamforming in a practical multiuser multitarget (MUMT) MIMO--OFDM system?
To answer this, we statistically characterize the discrete-periodic AF (DP-AF) and derive an exact closed-form expected sidelobe level (ESL).
We observe that nonzero-Doppler ESL cuts are delay-flat and the full two-dimensional integrated sidelobe level ratio (ISLR) is invariant to beamforming, whereas the zero-Doppler peak-to-sidelobe ratio (PSLR) remains controllable.
We accordingly formulate an AF-aware joint beamforming and subcarrier-allocation problem with target-wise PSLR and illumination constraints.
Since the direct PSLR constraint is intractable, we characterize its global optimum and prove that, for symbol kurtosis greater than one, the PSLR is maximized if and only if each target-wise directional-power profile is flat across subcarriers.
We then introduce the coefficient of variation (CV) to measure deviation from this flatness and derive tight CV-dependent PSLR bounds.
The lower bound yields a convex sufficient constraint for a desired PSLR, enabling an efficient reformulation.
Simulations show $9.4$--$11.0$ dB worst-case PSLR gains over communication-only beamforming at the flat-spectrum point with a $2.4$--$5.9\%$ sum-rate loss.
The CV-relaxed formulation achieves $99.7\%$ of the average sum-rate of the original PSLR-constrained problem, while reducing runtime by factors of $10.35$--$12.15$ under the considered scenarios.
\end{abstract}

\begin{IEEEkeywords}
Integrated sensing and communications (ISAC), dual-function radar-communications (DFRC), MIMO--OFDM, ambiguity function (AF), peak-to-sidelobe ratio (PSLR), OFDMA resource allocation, beamforming.
\end{IEEEkeywords}

\section{Introduction}

Integrated sensing and communications (ISAC) is envisioned to enable emerging applications such as autonomous mobility, smart manufacturing, and physical AI systems by allowing wireless networks to perceive their surrounding environments while providing communication services.
By sharing spectrum and hardware between sensing and communications, ISAC can efficiently integrate these two functions within a common wireless infrastructure \cite{LiuJSAC2022}.
Reflecting this potential, ISAC has been identified as a key usage scenario for IMT-2030 \cite{ITU_M2160_2023}.
To evolve existing cellular systems toward ISAC, 3GPP RAN1 has agreed to use cyclic-prefix OFDM (CP-OFDM) as the starting point for 6G ISAC waveform studies in Release 20 \cite{3GPP_R1_2603284}.
Combined with multiple-input multiple-output (MIMO) arrays, ISAC with CP-OFDM can provide a flexible time--frequency grid for delay--Doppler sensing and spatial degrees of freedom (DoF), enabling multiuser and multitarget (MUMT) operation by concurrently serving communication users and illuminating sensing targets.

\begin{figure*}[!t]
    \centering
    \includegraphics[width=\textwidth]{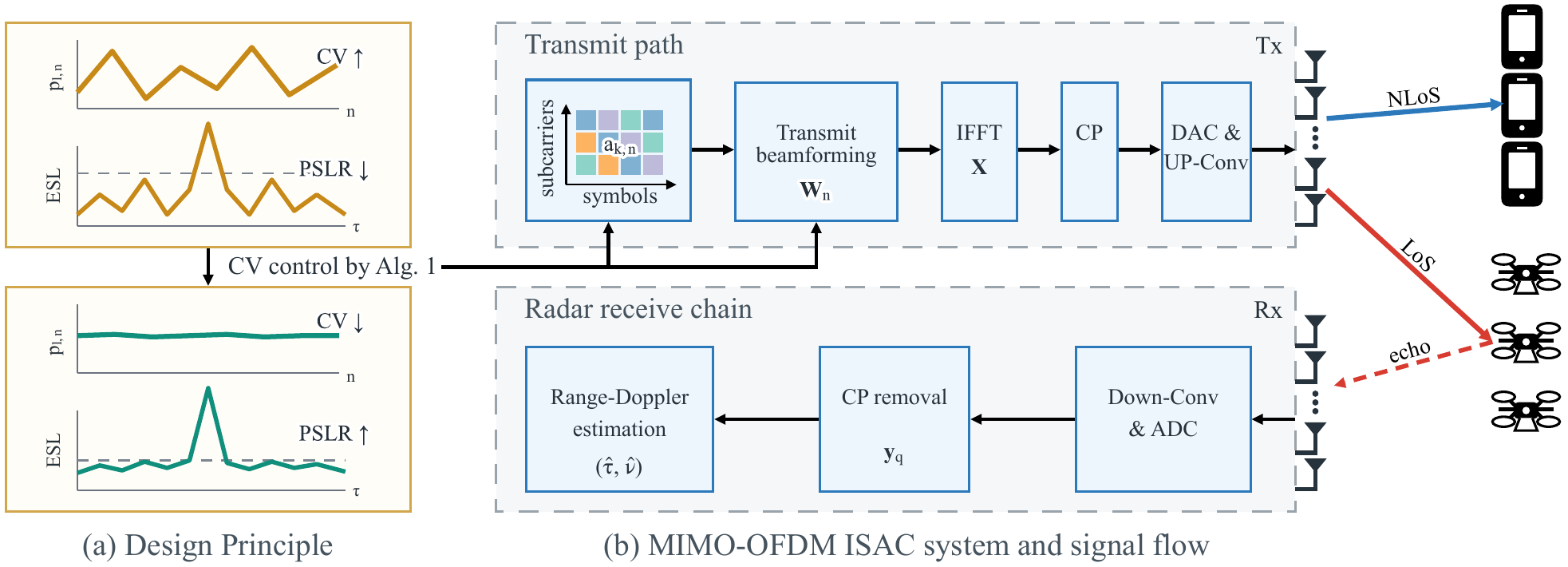}
    \caption{Proposed CV-based design and MIMO-OFDM ISAC signal flow.}
    \label{fig:system_model}
\end{figure*}

For MUMT sensing, the transmit waveform should be designed to resolve closely spaced targets and prevent strong target returns from obscuring weaker ones.
In radar systems, these waveform-dependent effects are classically characterized by Woodward's ambiguity function (AF), which describes the waveform response over delay--Doppler offsets \cite{Woodward1953,SturmProcIEEE2011}.
The AF provides a natural performance criterion for controlling the ambiguity and sidelobe structure that governs multitarget sensing performance, making AF shaping a fundamental principle of radar waveform design.

This deterministic AF-shaping, however, does not extend straightforwardly to OFDM ISAC. Unlike dedicated radar waveforms, OFDM ISAC signals are modulated by random communication symbols, so both the transmitted waveform and its instantaneous AF vary with the particular data realization. Consequently, even with fixed beamforming and resource allocation, the sidelobe pattern varies for each transmission, and unfavorable symbol realizations can produce pronounced range--Doppler sidelobes that obscure weak targets or generate spurious peaks \cite{li2025mimo,ZhangICASSP2026}. Thus, the AF in OFDM ISAC must be treated stochastically and data-dependently, motivating its statistical characterization and waveform optimization.

Most existing ISAC waveform schemes are instead formulated around sensing metrics such as receive signal-to-interference-plus-noise ratio (SINR), beampattern-matching error, mutual information (MI), and the Cram\'{e}r--Rao bound (CRB) \cite{WeiIoTJ2023,LuoCOMST2026}. These metrics provide tractable measures of complementary aspects of sensing performance, but they do not explicitly capture the statistical sidelobe behavior arising from random communication symbols. As a result, the ambiguity characteristics relevant to weak-target detection are not directly reflected in such objectives. This motivates a statistical characterization of the AF for ISAC OFDM waveforms and the incorporation of AF-based sidelobe control into MUMT ISAC waveform design.

Motivated by this need, recent studies have begun to characterize ISAC AFs or explicitly suppress their sidelobes. In particular, \cite{ZhangICASSP2026} rigorously characterized the statistical discrete-periodic AF (DP-AF) of random communication waveforms under arbitrary orthonormal modulation, deriving closed-form average sidelobe level and expected integrated sidelobe level (EISL) expressions.
This framework, however, does not address how the statistical DP-AF changes under spatial MIMO–OFDM precoding, where subcarrier-wise spatial transmission is jointly governed by MUMT requirements.

Other studies investigated AF shaping, but mostly under simplified systems (e.g., SISO, single-user and single-target, and dedicated sensing resources) and ad-hoc metrics (e.g. ISLR and PSLR) \cite{liao2025pulse,he2024dual,liu2025cp,ni2026ambiguity,liu2025beamforming}.
More relevant MIMO--OFDM works use symbol-level precoding (SLP) for instantaneous sidelobe suppression \cite{li2024low,li2025mimo,li2025lowcomplexity}.
However, SLP requires a high-dimensional per-block optimization, while existing formulations rely solely on the ISL for AF characterization and focus on a nominal target direction.
Consequently, a unified framework that characterizes the DP-AF in a MUMT MIMO--OFDM setting, reveals the effective AF metric from its structural properties, and integrates it into a joint communication--sensing optimization remains lacking.

To address these gaps, we establish a unified analytical framework for the DP-AF in a MUMT MIMO--OFDM ISAC system.
Using the ESL derived under this setting, we comparatively analyze three AF metrics---delay--Doppler coupling, ISLR, and PSLR---and show that only the PSLR retains nontrivial dependence on the transmit beamformer in a MUMT MIMO CP-OFDM system.
We further characterize the global optimum of the zero-Doppler PSLR and relate it to the cross-subcarrier flatness of each target-wise directional power profile. Then, we show that the PSLR is bounded by the coefficient-of-variation (CV) variable which is useful to convert the PSLR constraints into single convex sufficient conditions.
The overall CV-based design and the corresponding MIMO--OFDM ISAC signal flow are illustrated in Fig.~\ref{fig:system_model}.

The main contributions are summarized as threefold:
\begin{itemize}[leftmargin=*]
    \item We derive the DP-AF for a MUMT MIMO--OFDM ISAC system and obtain an exact closed-form expression for its ESL in terms of the subcarrier-wise directional transmit powers (Lemma~\ref{lem:ESL}). Using this expression, we analyze the three AF metrics according to their dependency on the transmit beamformer.
    Specifically, we establish that every nonzero-Doppler ESL cut is delay-flat, preventing beamformer-induced delay--Doppler coupling, and that the full two-dimensional ISLR remains strictly invariant at $N-1$ regardless of the beamformer configuration. 
    In contrast, the zero-Doppler PSLR varies with the subcarrier-wise directional powers and therefore provides the relevant AF metric. The first two properties extend the DP-AF results in \cite{ZhangICASSP2026} to the considered MIMO setting.
    \item Upon the above AF characterization, we formulate a novel joint transmit-beamforming and subcarrier-allocation problem~$(\mathcal{P}_{1})$ incorporating target-wise PSLR and minimum-illumination constraints. We further characterize the global maximum of the PSLR and prove that, for $\kappa>1$ and nonzero target illumination, it is achieved if and only if the target-wise directional transmit power is flat across the subcarriers (Theorem~\ref{thm: PSLR global bound}).
    \item Motivated by this optimality condition, we introduce the CV of the directional transmit powers and derive tight CV-dependent lower and upper bounds on the PSLR (Theorem~\ref{thm:cv_pslr_bound}). The lower bound yields a single sufficient convex constraint that replaces the direct PSLR constraint, making the proposed problem tractable. We then develop an efficient alternating-optimization algorithm that achieves $99.7\%$ of the direct PSLR-constrained baseline's average sum-rate while reducing wall-clock runtime by factors of $10.35$--$12.15$ under the considered stress scenarios (Figs.~\ref{fig:pareto_grid} and~\ref{fig:stress_burden}).
\end{itemize}

Section~II presents the system model and AF-based sensing metrics. Section~III develops the CV reformulation, Section~IV describes the optimization algorithms, and Section~V reports the numerical results. Section~VI concludes the paper.

\emph{Notations}: Matrices and vectors are denoted by bold uppercase and bold lowercase letters, respectively;
$\otimes$ denotes the Kronecker product,
$(\cdot)^*$, $(\cdot)^{\mathsf{T}}$, and $(\cdot)^{\mathsf{H}}$ are complex conjugate, transpose, and Hermitian transpose, respectively;
$\|\cdot\|_p$ and $\|\cdot\|_F$ stand for the $\ell_p$-norm and the Frobenius norm;
$\mathbb{E}[\cdot]$ is the expectation, $\mathrm{diag}(\cdot)$ is the operation of generating a diagonal matrix, and $\delta_{i,j}$ is the Kronecker delta function (i.e., $\delta_{i,j}=1$ for $i=j$, otherwise, $\delta_{i,j}=0$).

\section{System Model}

Consider a downlink MIMO--OFDM ISAC system in which the base station (BS) is equipped with colocated transmit and receive uniform linear arrays (ULAs), comprising $N_T$ and $N_R$ antenna elements with inter-element spacings $d_T$ and $d_R$, respectively. It uses the same data-bearing OFDM waveform to serve $K$ single-antenna users and illuminate $L$ far-field targets over $N$ subcarriers \cite{noh2023ici,noh2026dcfnet}. No dedicated radar waveform is superimposed; transmit beamforming controls the spatial response, while orthogonal frequency-division multiple access (OFDMA) assigns each subcarrier exclusively \cite{ChengJSTSP2021,LiuVTC2024}.

The communication links are modeled as rich-scattering non-line-of-sight (NLoS) channels with Rayleigh fading. The sensing targets are modeled as far-field point targets with line-of-sight (LoS) paths, and their azimuth angles are assumed known at the BS \cite{ChengTCCN2021}. Define the user, target, and subcarrier index sets as $\mathcal{K}\triangleq\{0,\ldots,K-1\}$, $\mathcal{L}\triangleq\{0,\ldots,L-1\}$, and $\mathcal{N}\triangleq\{0,\ldots,N-1\}$, respectively.

Let $\mathbf{s}=[s_0,s_1,\ldots,s_{N-1}]^{\mathsf{T}}\in\mathbb{C}^{N\times 1}$ denote the frequency-domain data-symbol vector. The symbols are drawn independently from the same zero-mean, unit-energy constellation and satisfy $\mathbb{E}[s_n]=\mathbb{E}[s_n^2]=0$ and $\mathbb{E}[|s_n|^2]=1$. Their fourth moment is denoted by $\kappa\triangleq\mathbb{E}[|s_n|^4]$; for example, $\kappa=1$ for constant-modulus PSK signaling such as QPSK, $\kappa\approx1.32$ for 16-QAM, and $\kappa=2$ for a complex Gaussian symbol \cite{liu2025uncovering}.

\subsection{Downlink Transmit Signal and Communication Model}
\subsubsection{Downlink MIMO--OFDM ISAC Transmit Signal}
Let $\mathbf{w}_n \in \mathbb{C}^{N_T\times1}$ denote the transmit beamforming vector applied to the data symbol on subcarrier $n$. The aggregated beamforming matrix across all subcarriers is defined as
\begin{equation}
    \mathbf{W} = \begin{bmatrix}
        \mathbf{w}_0 &
        \mathbf{w}_1 &
        \ldots &
        \mathbf{w}_{N-1}
    \end{bmatrix}^{\mathsf{T}} \in \mathbb{C}^{N\times N_T}.
\end{equation}
The discrete-time MIMO--OFDM ISAC transmit signal matrix is then compactly expressed as
\begin{equation} \label{eq:transmitted_ISAC_signal}
    \mathbf{X} = \mathbf{F}_N^{\mathsf{H}} \mathrm{diag}(\mathbf{s}) \mathbf{W},
\end{equation}
where $\mathbf{F}_N$ denotes the $N$-point unitary DFT matrix.
The transmit beamformers satisfy the BS power-budget constraint
\begin{equation} \label{eq:power_budget}
    \sum_{n\in\mathcal{N}} \|\mathbf{w}_n\|_2^2 \leq P_{\max}.
\end{equation}

\subsubsection{Downlink Communication Model}
Let $\mathbf{h}_{k,n} \in \mathbb{C}^{N_T\times 1}$ denote the frequency-domain Rayleigh fading channel vector from the BS to user $k$ on subcarrier $n$. 
Let $\alpha_{k,n} \in \{0, 1\}$ denote the binary subcarrier allocation indicator, where $\alpha_{k,n} = 1$ if subcarrier $n$ is allocated to user $k$, and $\alpha_{k,n} = 0$ otherwise.

If subcarrier $n$ is allocated to user $k$, the received baseband signal after CP removal is given by
\begin{equation}
    y_{k,n} = \mathbf{h}_{k,n}^{\mathsf{H}} \mathbf{w}_n s_n + z_{k,n},
\end{equation}
where $z_{k,n}\sim\mathcal{CN}(0,\sigma^2)$ is additive white Gaussian noise at user $k$. Under the unit-subcarrier-bandwidth normalization, the corresponding per-subcarrier spectral efficiency is
\begin{equation}
    \gamma_{k,n} = \log_2\left(1 + \frac{|\mathbf{h}^{\mathsf{H}}_{k,n} \mathbf{w}_n|^2}{\sigma^2}\right).
\end{equation}
Then, the resulting aggregate rate is
\begin{equation}\label{eq:aggregate_rate}
    \gamma_{\mathrm{tot}} = \sum_{k\in\mathcal{K}}\sum_{n\in\mathcal{N}} \alpha_{k,n}\gamma_{k,n}.
\end{equation}

\subsection{MIMO DP-AF}
\subsubsection{General MIMO DP-AF}
Consider a far-field target characterized by the parameter tuple $(\tau, \nu, \theta)$, where $\tau, \nu \in \mathcal{N}$ denote the integer delay and Doppler-shift indices, respectively, and $\theta$ denotes the azimuth angle.

Let $\mathbf{x}_p \in \mathbb{C}^{N \times 1}$ represent the discrete-time baseband signal transmitted from the $p$-th antenna element. The received echo signal at the $q$-th receive element is given by
\begin{equation}\label{eq:rcvd_echo}
    \mathbf{y}_q = \beta \sum_{p=0}^{N_T-1} \mathbf{D}_{N,\nu} \mathbf{J}_{N,\tau} \mathbf{x}_p e^{j2\pi\frac{(p d_T + q d_R)\sin\theta}{\lambda}},
\end{equation}
where $\beta$ is the complex channel gain and $\lambda$ is the carrier wavelength.
The matrices for periodic time-shift and frequency-shift operations are given by
\begin{equation}
    \mathbf{J}_{N,\tau} \!\triangleq\!
    \begin{bmatrix}
        \mathbf{0} & \mathbf{I}_{\tau}\\
        \mathbf{I}_{N-\tau} & \mathbf{0}
    \end{bmatrix}\!, \
    \!\!\mathbf{D}_{N,\nu} \!\triangleq\! \mathrm{diag}\Big(\!1, e^{j2\pi\frac{\nu}{N}}, \dots, e^{j2\pi\frac{(N-1)\nu}{N}}\!\Big).
\end{equation}

We assume that the maximum round-trip delay of the paths under consideration does not exceed the CP duration. After CP removal, the corresponding delay operation is represented by the periodic shift matrix $\mathbf{J}_{N,\tau}$ as in \eqref{eq:rcvd_echo}.
Accordingly, the following analysis considers the sampled periodic AF on the integer delay--Doppler grid $(\tau,\nu)\in\mathcal{N}^2$ after CP removal.

Assuming the target azimuth angle $\theta$ is known a priori, the matched filter output for hypothesized target parameters $(\tau', \nu', \theta)$ is formulated as
\begin{subequations}
\begin{align}
    &\sum_{q=0}^{N_R-1} \mathbf{y}_q^{(\tau',\nu',\theta)\mathsf{H}} \mathbf{y}_q^{(\tau,\nu,\theta)} \\
    &=\sum_{q=0}^{N_R-1} \bigg( \sum_{p'=0}^{N_T-1} \mathbf{D}_{N,\nu'} \mathbf{J}_{N,\tau'} \mathbf{x}_{p'} e^{j2\pi\frac{(p' d_T + q d_R)\sin\theta}{\lambda}} \bigg)^{\mathsf{H}} \\[-1mm]
    &\hspace{16mm} \times \bigg( \sum_{p=0}^{N_T-1} \mathbf{D}_{N,\nu} \mathbf{J}_{N,\tau} \mathbf{x}_p e^{j2\pi\frac{(p d_T + q d_R)\sin\theta}{\lambda}} \bigg)
    \nonumber\\
    &=N_R e^{-j2\pi\frac{\tau(\nu'-\nu)}{N}} \sum_{p, p'=0}^{N_T-1} e^{-j2\pi\frac{p' d_T \sin\theta}{\lambda}} \mathbf{x}_{p'}^{\mathsf{H}} \\[-2mm]
    &\hspace{36mm} \times \mathbf{J}_{N,\tau'-\tau}^{\mathsf{T}} \mathbf{D}_{N,\nu'-\nu}^* e^{j2\pi\frac{p d_T \sin\theta}{\lambda}} \mathbf{x}_p
    \nonumber\\
    &= N_R e^{-j2\pi\frac{\tau(\nu'-\nu)}{N}} \mathbf{a}^{\mathsf{H}}(\theta) \mathbf{X}^{\mathsf{H}} \mathbf{J}^{\mathsf{T}}_{N,\tau'-\tau} \mathbf{D}^*_{N,\nu'-\nu} \mathbf{X} \mathbf{a}(\theta),
\end{align}
\end{subequations}
where $\mathbf{a}(\theta) = \begin{bmatrix}
    1 & e^{j2\pi\frac{d_T\sin\theta}{\lambda}} & \cdots & e^{j2\pi\frac{(N_T-1)d_T\sin\theta}{\lambda}}
    \end{bmatrix}^{\mathsf{T}} \in \mathbb{C}^{N_T\times 1}$
is the transmit array steering vector. The receive array contributes the common factor $N_R$ in the matched-filter expression preceding \eqref{eq:General_Discrete_MIMO_AF}, which cancels from the normalized AF ratios.
Consequently, the general DP-AF for a MIMO ISAC system is defined as \cite{ChenVaidyanathanTSP2008,LiVorobyovTSP2015,ZhangICASSP2026}
\begin{equation} \label{eq:General_Discrete_MIMO_AF}
    \chi[\tau, \nu; \theta] = \mathbf{a}^{\mathsf{H}}(\theta) \mathbf{X}^{\mathsf{H}} \mathbf{J}_{N,\tau}^{\mathsf{T}} \mathbf{D}^*_{N,\nu} \mathbf{X} \mathbf{a}(\theta).
\end{equation}
\subsubsection{Sidelobe Level of AF}
Substituting our transmit ISAC signal \eqref{eq:transmitted_ISAC_signal} into the DP-AF \eqref{eq:General_Discrete_MIMO_AF}, the AF can be expressed as a function of the data symbols and beamforming vectors:
\begin{align}
    &\chi[\tau, \nu; \theta] \nonumber\\
    &\hspace{-1mm}=\! \mathbf{a}^{\mathsf{H}}(\theta) \mathbf{W}^{\mathsf{H}} \mathrm{diag}(\mathbf{s})^* \mathbf{F}_N \mathbf{J}^{\mathsf{T}}_{N,\tau} \mathbf{D}^*_{N,\nu} \mathbf{F}_N^{\mathsf{H}} \mathrm{diag}(\mathbf{s}) \mathbf{W} \mathbf{a}(\theta) \\
    &\hspace{-1mm}=\! \mathbf{s}^{\mathsf{H}} \mathrm{diag}(\mathbf{W}\mathbf{a}(\theta))^* \mathbf{F}_N \mathbf{J}^{\mathsf{T}}_{N,\tau} \mathbf{D}^*_{N,\nu} \mathbf{F}_N^{\mathsf{H}} \mathrm{diag}(\mathbf{W}\mathbf{a}(\theta)) \mathbf{s}.
\end{align}

Due to the stochastic nature of the communication symbols within the ISAC waveform, the AF exhibits random fluctuations. Therefore, the main and sidelobe levels must be evaluated probabilistically \cite{liu2025uncovering,ZhangICASSP2026}.
Once we define the \emph{per-subcarrier directional power} at angle $\theta$ as $P_n(\theta) \triangleq |\mathbf{w}_n^{\mathsf{T}} \mathbf{a}(\theta)|^2$, the Expected Sidelobe Level (ESL) can be expressed by the following lemma:

\begin{lemma}[Closed-form ESL] \label{lem:ESL}
Define the ESL as $\mathrm{ESL}[\tau,\nu;\theta] \triangleq \mathbb{E}[|\chi[\tau,\nu;\theta]|^2]$. The exact closed-form expression for the MIMO--OFDM ESL is given by
\begin{align} \label{eq:ESL_closed_form}
    \mathrm{ESL}[\tau,\nu;\theta]
    &= \delta_{\nu,0} \Big( \sum_{a\in\mathcal{N}}\sum_{b\in\mathcal{N}} P_a(\theta) P_b(\theta) e^{-j2\pi\frac{(a-b)\tau}{N}} \nonumber \\[-1mm]
    &\hspace{-7mm} + (\kappa-2)\sum_{a\in\mathcal{N}} P_a^2(\theta) \Big)
    + \sum_{a\in\mathcal{N}} P_a(\theta) P_{a-\nu}(\theta).
\end{align}
For the zero-Doppler cut (i.e., $\nu=0$), the ESL is given by
\begin{equation}
    \hspace{-2mm}\mathrm{ESL}[\tau,0;\theta]
    \hspace{-1mm}=\hspace{-1mm} (\kappa - 1) \sum_{n\in\mathcal{N}}\hspace{-.5mm} P_n(\theta)^2
    \hspace{-1mm}+\hspace{-1mm} \Big| \sum_{n\in\mathcal{N}}\hspace{-.5mm} P_n(\theta)e^\frac{j2\pi n\tau}{N} \Big|^2.
\end{equation}
\end{lemma}
\begin{proof}
    See Appendix~\ref{app:proof_ESL}.
\end{proof}

\subsection{AF-based Sensing Metrics}

An ideal AF should exhibit a sharp mainlobe accompanied by a low sidelobe floor. To quantitatively evaluate the AF shape, the following three characteristics are widely adopted:

\noindent\textbf{Delay--Doppler Coupling (DD-coupling):} Exemplified by the tilted AF ridge of linear frequency modulation (LFM) waveforms, DD-coupling causes Doppler-dependent range bias and degrades joint delay and Doppler resolution \cite{Woodward1953,NajianSebt2019}.

\vspace{0.1cm}
\noindent\textbf{Integrated Sidelobe Level Ratio (ISLR):} Defined as the ratio of the total sidelobe energy to the mainlobe energy, the ISLR provides an aggregate measure of AF sidelobes and is widely used as a waveform-shaping criterion \cite{AubryTSP2013,KeskinTSP2021,SongTSP2016,AlaeeTSP2017}.

\vspace{0.1cm}
\noindent\textbf{Peak-to-Sidelobe Ratio (PSLR):} On the zero-Doppler ESL cut, PSLR is the mainlobe-to-peak-sidelobe ratio and an indicator of masking susceptibility \cite{KeskinEuRAD2021,SongTSP2016,AlaeeTSP2017}.

\vspace{0.1cm}
Rather than selecting a single AF sidelobe metric in advance, we analyze all three metrics above using our closed-form ESL in Lemma~\ref{lem:ESL}. The analysis shows that DD-coupling is absent and the full two-dimensional ISLR is invariant, consistent with the scalar OFDM waveform results of \cite{ZhangICASSP2026} and demonstrating that these properties persist under our direction-dependent MIMO beamforming extension. In contrast, the zero-Doppler PSLR retains design freedom.

\subsubsection{DD-coupling Analysis}
The delay-dependent term in the ESL contributes only on the zero-Doppler cut by the Kronecker delta $\delta_{\nu,0}$. For any fixed non-zero Doppler cut $\nu \neq 0$, the ESL reduces to
\begin{equation}\label{eq:nonzero_doppler_esl}
    \mathrm{ESL}[\tau,\nu;\theta] = \sum_{a\in\mathcal{N}} P_a(\theta) P_{a-\nu}(\theta),
\end{equation}
Thus, every nonzero-Doppler ESL cut is delay-flat and exhibits no delay--Doppler coupling. The same Kronecker-delta structure appears in the scalar OFDM DP-AF~\cite{ZhangICASSP2026}, and \eqref{eq:nonzero_doppler_esl} shows that it persists under directional MIMO transmit beamforming.

\subsubsection{Closed-form Derivation of ISLR}

We define the ISLR at the target angle $\theta$ as
\begin{align}
    \mathrm{ISLR}(\theta)\!\triangleq\!\frac{\sum_{\tau\in\mathcal{N}}\sum_{\nu\in\mathcal{N}}\mathrm{ESL}[\tau,\nu;\theta]-\mathrm{ESL}[0,0;\theta]}{\mathrm{ESL}[0,0;\theta]}.
\end{align}
Let $Q(\theta)\triangleq\sum_{n\in\mathcal{N}}P_n(\theta)^2$ and $S(\theta)\triangleq\sum_{n\in\mathcal{N}}P_n(\theta)$.
Then, the total AF energy and the expected mainlobe level can be obtained from Lemma~\ref{lem:ESL}, yielding\footnote{
The derivation of \eqref{eq:AF_total_energy} is given in Appendix~\ref{app:deriv_ISL}.
}
\begin{alignat}{2}
    &\mathrm{ESL}[0,0;\theta]&&=(\kappa-1)Q(\theta)+S(\theta)^2,
    \\
    \sum_{\tau\in\mathcal{N}}\sum_{\nu\in\mathcal{N}}&\mathrm{ESL}[\tau,\nu;\theta]
    &&=N\big[(\kappa-1)Q(\theta)+S(\theta)^2\big] \nonumber\\[-3mm]
    & &&=N\,\mathrm{ESL}[0,0;\theta].
    \label{eq:AF_total_energy}
\end{alignat}
Therefore, the ISLR at the target angle $\theta$ reduces to a constant,
\begin{equation}\label{eq:islr_constant}
    \mathrm{ISLR}(\theta)=N-1.
\end{equation}
This invariance reflects Woodward's classical AF volume property and was recently shown to persist for the normalized EISL of scalar ISAC waveforms, including OFDM \cite{Woodward1953,ZhangICASSP2026}. Equations~\eqref{eq:AF_total_energy} and~\eqref{eq:islr_constant} extend this result to direction-dependent MIMO transmit beamforming, for which the ISLR remains fixed at $N-1$ and is therefore omitted from the optimization considerations.

\subsubsection{Closed-form Derivation of the Zero-Doppler PSLR}
After Doppler compensation, we evaluate range sidelobes on the zero-Doppler ESL cut, since Lemma~\ref{lem:ESL} shows that the ESL at any nonzero Doppler is independent of delay and thus contains no range-sidelobe structure. This treatment is consistent with OFDM ranging analyses based on the autocorrelation function (ACF) and peak sidelobe level (PSL)~\cite{liu2025uncovering,KeskinEuRAD2021}.

We define the PSLR at the target angle $\theta$ for the zero-Doppler cut as
\begin{subequations}\label{eq:pslr_definition}
\begin{align}
    &\mathrm{PSLR}(\theta) \triangleq
    \frac{\mathrm{ESL}[0,0;\theta]}{\max_{\tau\in\mathcal{N}\setminus\{0\}}\mathrm{ESL}[\tau,0;\theta]} \\
    &=\frac{(\kappa-1)Q(\theta)+S(\theta)^2}{(\kappa-1)Q(\theta)+\max\limits_{\tau\in\mathcal{N}\setminus\{0\}}\left|\sum_{n\in\mathcal{N}}P_n(\theta)e^{\frac{j2\pi n\tau}{N}}\right|^2}.
\end{align}
\end{subequations}
Accordingly, the bounds and optimization claims of next section apply to this zero-Doppler range PSLR.

\subsection{Problem Formulation}

The preceding subsection analyzed the three AF characteristics using the closed-form ESL in Lemma~\ref{lem:ESL} and showed that, in the considered MIMO--OFDM setting with transmit beamforming, the zero-Doppler PSLR is the relevant controllable metric for AF shaping. Based on this result, we propose a novel joint beamforming and subcarrier-allocation formulation with explicit PSLR constraints.

For target $l\in\mathcal{L}$, define
\begin{align}
    \mathbf{A}_l &\triangleq \mathbf{a}^*(\theta_l)\mathbf{a}^{\mathsf{T}}(\theta_l),\\
    p_{l,n} &\triangleq P_n(\theta_l)
    =\operatorname{Tr}(\mathbf{A}_l\mathbf{W}_n)
    =\mathbf{a}^{\mathsf{T}}(\theta_l)\mathbf{W}_n\mathbf{a}^*(\theta_l),
    \label{eq:target_directional_power}
\end{align}
and collect the directional powers across subcarriers as
\begin{equation}\label{eq:target_directional_power_vector}
    \mathbf{p}_l \triangleq [p_{l,0},\ldots,p_{l,N-1}]^{\mathsf{T}}.
\end{equation}
Define the target-wise delay-domain response and mean directional power as
\begin{align}
    &R_{l,\tau}
    \triangleq R_\tau(\theta_l)
    =\sum_{n\in\mathcal{N}}p_{l,n}e^{j2\pi n\tau/N},~
    R_{l,0}=\mathbf{1}^{\mathsf{T}}\mathbf{p}_l,\\
    &\mu_l
    \triangleq \frac{1}{N}\sum_{n\in\mathcal{N}}p_{l,n}
      =\frac{1}{N}\mathbf{1}^{\mathsf{T}}\mathbf{p}_l
      =\frac{R_{l,0}}{N}.
\end{align}
The parameters $\gamma_{k,\min}$ and $P_{\mathrm{des},l}$ denote the minimum aggregate user rate normalized by the subcarrier spacing and the minimum transmit-side directional-power floor toward the $l$-th target direction, respectively. For brevity of notations, we denote $\mathrm{PSLR}_l\triangleq\mathrm{PSLR}(\theta_l)$. The proposed joint design problem is formulated as
\begin{subequations} \label{prob:p1}
\begin{alignat}{2}
    \hspace{-3mm}(\mathcal{P}_1): &\max_{\{\mathbf{W}_n \succeq 0\}, \{\alpha_{k,n}\}} 
    ~ && \gamma_{\mathrm{tot}} \label{const:p1_obj}\\
    &~~~~~~~\text{s.t.} ~ 
    && \mathrm{PSLR}_l \geq \mathrm{PSLR}_{\min}, ~ \forall l\in\mathcal{L}, \label{const:p1_pslr} \\
    & && \mu_l \geq P_{\mathrm{des},l}, ~ \forall l\in\mathcal{L}, \label{const:p1_illumination} \\
    & && \sum_{n\in\mathcal{N}} \alpha_{k,n}\gamma_{k,n} \geq \gamma_{k,\min}, ~ \forall k\in\mathcal{K}, \label{const:p1_qos} \\
    & && \sum_{n\in\mathcal{N}} \operatorname{Tr}(\mathbf{W}_n) \leq P_{\max}, \label{const:p1_power} \\
    & && \operatorname{rank}(\mathbf{W}_n)=1, ~ \forall n\in\mathcal{N}, \label{const:p1_rank} \\
    & && \sum_{k\in\mathcal{K}} \alpha_{k,n} \leq 1, ~ \forall n\in\mathcal{N}, \label{const:p1_exclusivity} \\
    & && \alpha_{k,n} \in \{0,1\}, ~ \forall k\in\mathcal{K},\ n\in\mathcal{N}. \label{const:p1_binary}
\end{alignat}
\end{subequations}

The PSLR constraint \eqref{const:p1_pslr} is accompanied by the illumination floor \eqref{const:p1_illumination}, which enforces a minimum transmit power toward each target and provides a guarantee for the minimum received echo SINR. It also prevents a degenerate solution in which the target illumination collapses to zero while shaping the scale-invariant PSLR constraint.

\section{PSLR Optimality and CV-Based Reformulation}
\subsection{Intractability of the PSLR Constraint}

Problem $(\mathcal{P}_1)$ is a mixed-integer nonlinear program (MINLP), for which we employ alternating optimization (AO) to update the subcarrier allocation and beamforming variables in turn \cite{RazaviyaynSIOPT2013}.
Even after this separation, the resulting beamforming subproblem remains intractable due to the computationally demanding PSLR constraint \eqref{const:p1_pslr}.

Recall that $\mathbf{p}_l=[p_{l,0},\ldots,p_{l,N-1}]^{\mathsf{T}}$ and $p_{l,n}=\operatorname{Tr}(\mathbf{A}_l\mathbf{W}_n)$, so $\mathbf{p}_l$ is affine in the beamforming covariance matrices $\{\mathbf{W}_n\}$. The PSLR constraint \eqref{const:p1_pslr} can be equivalently rewritten as a \emph{difference-of-convex (DC)} inequality:
\begin{equation} \label{eq:dc_pslr}
    \mathrm{PSLR}_{\min}\mathcal{D}_{\mathrm{P}}(\mathbf{p}_l)
    -\mathcal{N}_{\mathrm{P}}(\mathbf{p}_l) \leq 0,
\end{equation}
where the mainlobe and peak sidelobe terms are, respectively,
\begin{align}
    \mathcal{N}_{\mathrm{P}}(\mathbf{p}_l)
    &\triangleq |R_{l,0}|^2
      +(\kappa-1)\|\mathbf{p}_l\|_2^2,\\
    \mathcal{D}_{\mathrm{P}}(\mathbf{p}_l)
    &\triangleq \max_{\tau\ne0}|R_{l,\tau}|^2
      +(\kappa-1)\|\mathbf{p}_l\|_2^2,
\end{align}
where two terms are convex in $\mathbf{p}_l$.

Alternatively, the max operator can be eliminated by imposing the PSLR condition for every sidelobe, yielding \(L(N-1)\) SOC constraints. This increases the number of constraints in each AO beamforming SDP after semidefinite relaxation (SDR) from \(2L+K\) to \(L(N-1)+L+K\), substantially increasing the complexity.

To avoid such large SOC expansion, we retain the compact DC form and handle it using an inner successive convex approximation (SCA) loop \cite{ScutariTSP2017}, solving a convex conic subproblem at each iteration. This repeated solution still remains computationally costly, motivating the following compact CV-based reformulation.

\subsection{Flat Directional Power for Optimal AF Shaping}
Given the intractability of explicitly imposing the PSLR constraint \eqref{const:p1_pslr}, a practical approach is to identify a mathematically tractable alternative metric. As a fundamental step toward formulating this alternative, we evaluate the theoretical limit of the PSLR.

First, we obtain the theoretically optimal PSLR in the following theorem:

\begin{theorem}[Global Maximum PSLR]\label{thm: PSLR global bound}
For $\kappa > 1$ and $\mathbf{1}^{\mathsf{T}}\mathbf{p}_l>0$, the $\mathrm{PSLR}_l$ achieves its global maximum if and only if the directional power is flat across all $N$ subcarriers, i.e., $p_{l,0} = p_{l,1} = \cdots = p_{l,N-1}$. Accordingly, the globally optimal PSLR over the directional-power profiles is given by
\begin{equation}
    \mathrm{PSLR}_{l,\max} = 1 + \frac{N}{\kappa - 1}.
    \label{eq:pslr_max}
\end{equation}
\end{theorem}
\begin{proof}
    See Appendix~\ref{app:proof_thm1}.
\end{proof}

Theorem~\ref{thm: PSLR global bound} shows that, for $\kappa>1$, the PSLR is globally maximized if and only if the directional power is \textit{perfectly flat across subcarriers}. The corresponding maximum is $1+N/(\kappa-1)$. In the special case $\kappa=1$, the common sidelobe-floor term vanishes, and the PSLR has no finite upper bound.

Although exact flatness maximizes the PSLR, it may impose a stricter PSLR requirement than necessary for the intended sensing operation, thereby reducing the beamforming freedom available for communication.
We therefore introduce a tunable relaxation of the flatness condition to enable a flexible sensing--communication trade-off.

The structural requirement of the optimal PSLR condition suggests that we can regulate the PSLR by penalizing the variation of $\{p_{l,n}\}_{n\in\mathcal{N}}$. For target $l\in\mathcal{L}$, we quantify this variation by the coefficient of variation (CV), defined as
\begin{equation}\label{eq:cv_definition}
    \sigma_l^2\triangleq\frac{1}{N}\|\mathbf{p}_l\|_2^2-\mu_l^2,
    \qquad
    \mathrm{CV}_l\triangleq\frac{\sigma_l}{\mu_l}.
\end{equation}

Given a specific CV, the PSLR is theoretically bounded as described in the following theorem:
\begin{theorem}[CV-Based PSLR Bound]\label{thm:cv_pslr_bound}

For each target $l\in\mathcal{L}$ with $\mathbf{1}^{\mathsf{T}}\mathbf{p}_l>0$, the upper bound below holds for any $N$. For even $N$, the PSLR satisfies the following two-sided bound:
\begin{equation}
    \mathcal{L}(N, \kappa, \mathrm{CV}_l) \leq \mathrm{PSLR}_l \leq \mathcal{U}(N, \kappa, \mathrm{CV}_l).
\end{equation}
where $\mathcal{U}(N, \kappa, \mathrm{CV}_l)$ and $\mathcal{L}(N, \kappa, \mathrm{CV}_l)$ are given by
\begin{align}
    &\mathcal{U}(N, \kappa, \mathrm{CV}_l) \!=\! \frac{\kappa - 1}{\kappa - 1 + \frac{N}{N-1}} \!+\! \frac{\frac{N^2\kappa}{(N-1)\left(\kappa - 1 + \frac{N}{N-1}\right)}}{\left(\kappa \!- \!1 \!+\! \frac{N}{N-1}\!\right)\mathrm{CV}_l^2 \!+\! \kappa \!-\! 1}, \label{eq:upper_bound} \\
    &\mathcal{L}(N, \kappa, \mathrm{CV}_l) \!=\! \frac{\kappa - 1}{N + \kappa - 1} \!+\! \frac{\frac{N(N + 2\kappa - 2)}{N + \kappa - 1}}{(N + \kappa - 1)\mathrm{CV}_l^2 + \kappa - 1}. \label{eq:lower_bound}
\end{align}

For $\kappa=1$ with $\mathrm{CV}_l>0$, the result simplifies to $1/\mathrm{CV}_l^2\leq\mathrm{PSLR}_l\leq(N-1)/\mathrm{CV}_l^2$.

\end{theorem}
\begin{proof}
    See Appendix~\ref{app:proof_thm3}.
\end{proof}
Fig.~\ref{fig:cv_pslr_theory} compares the analytic PSLR bounds with $100$ independent channel realizations. The shaded band spans the 5th to 95th percentiles across the simulations. Note that both bounds are strictly decreasing with respect to $\mathrm{CV}_l$. Thus, constraining the target-wise CV certifies the PSLR defined in \eqref{eq:pslr_definition} without directly optimizing its peak term.

\begin{figure}[!t]
    \centering
    \begingroup
    \setlength{\unitlength}{\columnwidth}
    \begin{picture}(1,0.53846)
        \put(0,0){\includegraphics[width=\columnwidth]{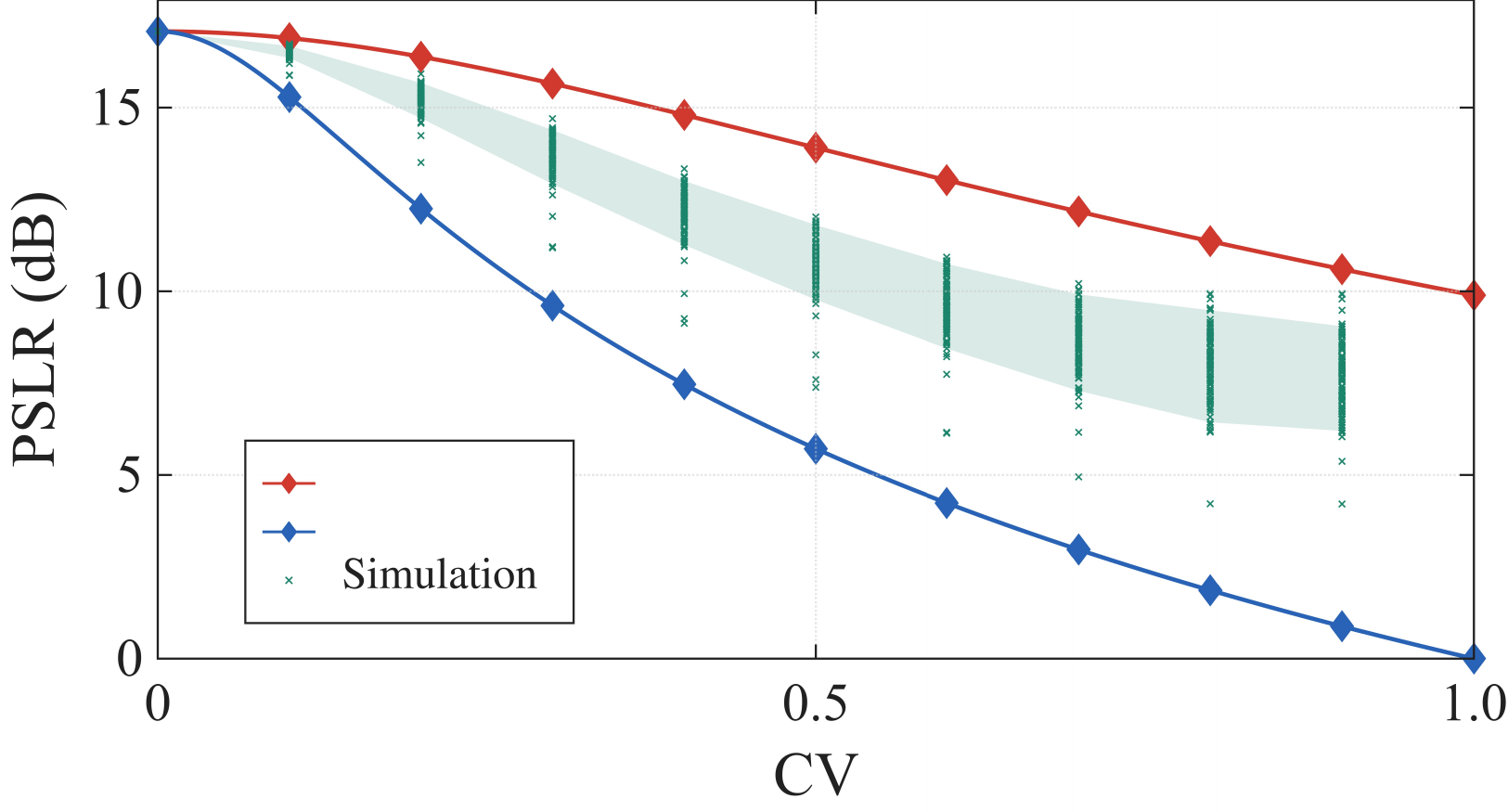}}
        \put(0.2305,0.2125){\makebox(0,0)[l]{\scriptsize$\mathcal{U}(\mathrm{CV})$}}
        \put(0.2305,0.1814){\makebox(0,0)[l]{\scriptsize$\mathcal{L}(\mathrm{CV})$}}
    \end{picture}
    \endgroup
    \vspace{-1.5em}
    \caption{Analytic PSLR bounds and simulated range versus CV.}
    \label{fig:cv_pslr_theory}
\end{figure}

Building upon these properties, we translate the PSLR constraint \eqref{const:p1_pslr} into a tractable CV constraint. To guarantee $\mathrm{PSLR}_l\geq\mathrm{PSLR}_{\min}$, we restrict $\mathrm{CV}_l$ such that $\mathcal{L}(N,\kappa,\mathrm{CV}_l)\geq\mathrm{PSLR}_{\min}$. Accordingly, we define
\begin{equation}
\mathrm{CV}_{\max}=\mathcal{L}^{-1}(\mathrm{PSLR}_{\min}).
\end{equation}

Bounding this CV provides an optimization-friendly approach, because the condition can be written as a single second-order cone (SOC) constraint
\begin{equation}
    \|\mathbf{p}_l-\mu_l\mathbf{1}\|_2
    \leq \sqrt{N}\mathrm{CV}_{\max}\mu_l,
    ~ \forall l\in\mathcal{L}.
\end{equation}
Since $\mathbf{p}_l$ is affine in $\{\mathbf{W}_n\}$, this is a convex constraint on the covariance matrices.

\subsection{Proposed CV-Based Reformulation}\label{sec:p2_reformulation}

By replacing the PSLR threshold \eqref{const:p1_pslr} with the CV upper bound using \eqref{eq:lower_bound}, $(\mathcal{P}_1)$ can be reformulated as
\begin{subequations}\label{prob:p_prop_full}
\begin{align}
    (\mathcal{P}_{prop}): &\max_{\{\mathbf{W}_n \succeq 0\}, \{\alpha_{k,n}\}} ~ \gamma_{\mathrm{tot}} \label{const:p2f_obj}\\
    \text{s.t.} ~ &
    \mathrm{CV}_l \!\leq\! \mathrm{CV}_{\max}, ~ \forall l\in\mathcal{L}, \label{const:p2f_cv} \\
    & \mu_l \geq P_{\mathrm{des},l}, ~ \forall l\in\mathcal{L}, \label{const:p2f_illum} \\
    & \sum_{n\in\mathcal{N}} \alpha_{k,n}\gamma_{k,n} \geq \gamma_{k,\min}, ~ \forall k\in\mathcal{K}, \label{const:p2f_qos} \\
    & \sum_{n\in\mathcal{N}} \operatorname{Tr}(\mathbf{W}_n) \leq P_{\max}, \label{const:p2f_power} \\
    & \operatorname{rank}(\mathbf{W}_n)=1, ~ \forall n\in\mathcal{N}, \label{const:p2f_rank} \\
    & \sum_{k\in\mathcal{K}} \alpha_{k,n} \leq 1, ~ \forall n\in\mathcal{N}, \label{const:p2f_excl} \\
    & \alpha_{k,n} \in \{0,1\}, ~ \forall k\in\mathcal{K},\ n\in\mathcal{N}. \label{const:p2f_binary}
\end{align}
\end{subequations}
Two non-convexities remain in $(\mathcal{P}_{prop})$: the binary allocation of $\{\alpha_{k,n}\}$ \eqref{const:p2f_binary}, which is addressed by an AO update, and the rank-one constraints \eqref{const:p2f_rank}, which are relaxed for each fixed allocation and subsequently handled by rank-one recovery.

Two distinct mechanisms can degrade the sum-rate achievable by $(\mathcal{P}_{prop})$.

\emph{Impact of the illumination floor.} Retained from $(\mathcal{P}_1)$, the illumination constraint \eqref{const:p2f_illum} enforces a transmit-side directional-illumination proxy by reserving power and spatial DoF toward the target directions, as in sensing-oriented dual-function radar-communications beamforming \cite{ChengJSTSP2021,noh2026star}. Its impact is evaluated in Section~\ref{subsec:illumination_constraint}.

\emph{Conservatism of the CV Surrogate for PSLR.} To guarantee $\mathrm{PSLR}_l\geq\mathrm{PSLR}_{\min}$, the proposed formulation imposes the sufficient condition $\mathrm{CV}_l\leq\mathrm{CV}_{\max}=\mathcal{L}^{-1}(\mathrm{PSLR}_{\min})$. The uncertainty of this PSLR certification is bounded by
\begin{equation}\label{eq:pslr_loss_bound}
\Delta_{\mathrm{PSLR}} \triangleq \mathcal{U}(N,\kappa,\mathrm{CV}_{\max}) - \mathcal{L}(N,\kappa,\mathrm{CV}_{\max}).
\end{equation}
Thus, $\Delta_{\mathrm{PSLR}}$ gives the width of the PSLR bracket at the selected threshold. The resulting performance degradation is evaluated in Section~\ref{subsec:pslr_pareto}.

\section{Proposed Joint Design Algorithms}
\label{sec:algorithms}

As discussed in Sec. III, the joint optimization involves coupled beamforming matrices $\{\mathbf{W}_n\}$ and discrete subcarrier allocation indicators $\{\alpha_{k,n}\}$. To tackle this, we employ an Alternating Optimization (AO) framework. In this section, we first introduce a Lagrangian-dual subgradient update, a standard decomposition technique for OFDMA resource allocation, followed by the specific beamforming optimization algorithms for both the proposed CV-constrained formulation and the direct PSLR-constrained baseline.

\subsection{Subcarrier Allocation via Lagrangian-Dual Decomposition}
For fixed beamforming matrices $\{\mathbf{W}_n\}$, we relax the binary constraint to $\alpha_{k,n} \in [0, 1]$ and apply a Lagrangian-dual subgradient update \cite{YuLuiTCOM2006,LiuVTC2024,yang2025joint}. We denote the quality of service (QoS) multipliers by $\bm{\zeta} = [\zeta_0, \dots, \zeta_{K-1}]^{\mathsf{T}} \succeq \mathbf{0}$. The partial Lagrangian over the allocation variables $\{\alpha_{k,n}\}$ and these multipliers is given by
\begin{equation}
    \mathcal{F}(\{\alpha_{k,n}\},\bm{\zeta})
    =\gamma_{\mathrm{tot}}
    +\sum_{k\in\mathcal{K}}\zeta_k
    \Big(\sum_{n\in\mathcal{N}}\alpha_{k,n}\gamma_{k,n}
    -\gamma_{k,\min}\Big).
\end{equation}
For a given set of multipliers $\bm{\zeta}$, the optimal allocation is decoupled across subcarriers, yielding a greedy assignment:
\begin{equation}\label{eq:greedy_alloc}
    \alpha_{k^*,n}=1, ~ \text{if } k^*=\arg\max_{k\in\mathcal{K}}(1+\zeta_k)\gamma_{k,n},
\end{equation}
and $\alpha_{k,n} = 0$ for all $k \neq k^*$. 

At dual iteration $i$, define the allocated user rate as $R_k^{(i)}\triangleq\sum_{n\in\mathcal{N}}\alpha_{k,n}^{(i)}\gamma_{k,n}$. The multipliers are updated by the dual-subgradient step
\begin{equation}\label{eq:dual_update}
    \zeta_k^{(i+1)}=\left[\zeta_k^{(i)}+\delta^{(i)}
    \left(\gamma_{k,\min}-R_k^{(i)}\right)\right]^+,
\end{equation}
where $\delta^{(i)}>0$ is the step size at the $i$-th iteration. A greedy repair then reduces any QoS shortfalls, and feasibility is checked after rank-one recovery.

\subsection{Proposed CV-Constrained Beamforming Optimization}
For a fixed allocation $\{\alpha_{k,n}\}$, the beamforming update is the following SDR of $(\mathcal{P}_{prop})$:
\begin{subequations}\label{prob:p_sdp}
\begin{alignat}{2}
    (\mathcal{P}_{\mathrm{SDP}}):~ &\max_{\{\mathbf{W}_n \succeq 0\}}
    && \gamma_{\mathrm{tot}} \label{const:p_sdp_obj}\\
    &~~~~\text{s.t.}
    && \eqref{const:p2f_cv},\ \eqref{const:p2f_illum},\ \eqref{const:p2f_qos},\ \text{and}\ \eqref{const:p2f_power}.
\end{alignat}
\end{subequations}

Because the CV constraint is SOC-representable, $(\mathcal{P}_{\mathrm{SDP}})$ is a convex conic program solvable by an interior-point solver. Rank-one beamformers are recovered by dominant-eigenvector extraction, with Gaussian randomization (GR) used if needed.

\begin{algorithm}[t]
\caption{Proposed CV-Constrained AO Algorithm}
\label{alg:proposed_ao}
\begin{algorithmic}[1]
\STATE \textbf{Input:} $\mathcal D\triangleq\{\{\mathbf h_{k,n}\},\{\theta_l\},\{\gamma_{k,\min}\},\{P_{\mathrm{des},l}\}\}$.
\STATE \hspace{\algorithmicindent} $P_{\max}$, $\mathrm{CV}_{\max}$, $\epsilon$, $\epsilon_{\mathrm{dual}}$, $t_{\max}$, $i_{\max}$.
\STATE $(\bm\alpha^{(0)},\mathbf W^{(0)})\leftarrow\operatorname{FeasibleInit}(\mathcal D)$; $(t,\Gamma_{\mathrm{best}})\leftarrow(0,-\infty)$.
\REPEAT
    \STATE $t\leftarrow t+1$.
    \STATE \textbf{Step 1: Beamforming Optimization (SDP)}
    \STATE $\mathbf W^{\mathrm{SDP}}\leftarrow\operatorname{Solve}\!\left((\mathcal P_{\mathrm{SDP}})\mid\bm\alpha^{(t-1)}\right)$.
    \STATE $\mathbf w^{(t)}\leftarrow\operatorname{Recover}_{\mathrm{EVD/GR}}(\mathbf W^{\mathrm{SDP}})$; $\bm\gamma^{(t)}\leftarrow\bm\gamma(\mathbf w^{(t)})$.
    \STATE $\mathbf W_n^{(t)}\leftarrow\mathbf w_n^{(t)}(\mathbf w_n^{(t)})^{\mathsf H}$, $\forall n\in\mathcal N$.
    \STATE $\Gamma^{(t)}\leftarrow\sum_{k,n}\alpha_{k,n}^{(t-1)}\gamma_{k,n}^{(t)}$.
    \IF{$\Gamma^{(t)}>\Gamma_{\mathrm{best}}$}
        \STATE $\bigl(\mathbf w^\star,\bm\alpha^\star,\Gamma_{\mathrm{best}}\bigr)\leftarrow\bigl(\mathbf w^{(t)},\bm\alpha^{(t-1)},\Gamma^{(t)}\bigr)$.
    \ENDIF
    \STATE \textbf{Step 2: Subcarrier Allocation (Dual Method)}
    \STATE $(\bm\zeta^{(0)},i)\leftarrow(\mathbf0,0)$.
    \REPEAT
        \STATE $k_n^{(i)}\leftarrow\arg\max_{k\in\mathcal K}(1+\zeta_k^{(i)})\gamma_{k,n}^{(t)}$; $\alpha_{k,n}^{(i)}\leftarrow\mathbf1_{\{k=k_n^{(i)}\}}$.
        \STATE $R_k^{(i)}\!\!\!\leftarrow\!\!\!\sum_n\!\alpha_{k,n}^{(i)}\gamma_{k,n}^{(t)}$; $\zeta_k^{(i+1)}\!\!\leftarrow\!\![\zeta_k^{(i)}\!+\!\delta^{(i)}(\gamma_{k,\min}\!-\!R_k^{(i)})]^+$.
        \STATE $i\leftarrow i+1$.
    \UNTIL{$|\mathcal F^{(i)}-\mathcal F^{(i-1)}|<\epsilon_{\mathrm{dual}}$ or $i=i_{\max}$.}
    \STATE $\bm\alpha^{(t)}\leftarrow\operatorname{Repair}_{\mathrm{QoS}}\!\left(\bm\alpha^{(i)},\bm\gamma^{(t)}\right)$.
\UNTIL{$|\Gamma^{(t)}-\Gamma^{(t-1)}|<\epsilon$ or $t=t_{\max}$.}
\STATE \textbf{Output:} $(\mathbf w^\star,\bm\alpha^\star)$.
\end{algorithmic}
\end{algorithm}

\subsection{Direct PSLR-Constrained Baseline}
\label{subsec:direct_pslr_baseline}
To evaluate the effectiveness of the proposed CV approach, we use the direct PSLR-constrained baseline. As discussed in Sec. III-A, to avoid increasing the number of PSLR constraints by a factor of \(N-1\), we retain the compact difference-of-convex (DC) representation of the direct PSLR constraint.

To handle this, we resort to the SCA method \cite{ScutariTSP2017}. Let $\mathbf{p}_l^{(m)}$ denote the directional power profile of target $l$ at the $m$-th inner SCA iteration. We linearize the concave component $-\mathcal{N}_{\mathrm{P}}(\mathbf{p}_l)$ in \eqref{eq:dc_pslr} using its first-order Taylor expansion:
\begin{equation}
    \tilde{\mathcal{N}}_{\mathrm{P}}(\mathbf{p}_l;\mathbf{p}_l^{(m)})
    =\mathcal{N}_{\mathrm{P}}(\mathbf{p}_l^{(m)})
    +\nabla\mathcal{N}_{\mathrm{P}}(\mathbf{p}_l^{(m)})^{\mathsf{T}}
    (\mathbf{p}_l-\mathbf{p}_l^{(m)}).
\end{equation}
The PSLR constraint is thus approximated by the single convex surrogate:
\begin{equation}
    \mathrm{PSLR}_{\min}\mathcal{D}_{\mathrm{P}}(\mathbf{p}_l)
    -\tilde{\mathcal{N}}_{\mathrm{P}}(\mathbf{p}_l;\mathbf{p}_l^{(m)})\leq0.
\end{equation}
Consequently, solving the beamforming subproblem of the direct PSLR-constrained baseline requires an \emph{inner iterative loop}. At each step $m$, a convexified SDP is solved using the current linearization point to update $\mathbf{p}^{(m+1)}$. This additional SCA loop increases the computational complexity compared with a single SDP solve in each beamforming update of the proposed CV-based algorithm. The same eigenvalue decomposition (EVD)/GR recovery is applied to its final SDR solution.

\section{Simulations and Numerical Results}

\subsection{Simulation Setup}
We evaluate a downlink MIMO--OFDM ISAC system in which a base station serves multiple single-antenna users while illuminating multiple target directions. Unless otherwise stated, the simulation parameters are summarized in Table~\ref{tab:sim_params}. The channels are frequency-domain Rayleigh fading channels, and the transmit array is a half-wavelength ULA.

\begin{table}[!t]
\centering
\caption{Simulation Parameters}
\label{tab:sim_params}
\begin{tabular}{l|l}
\hline
Parameter & Value \\
\hline
Carrier frequency & $3.5$ GHz \\
Transmit array & ULA, $N_T=4$, $d=\lambda/2$ \\
OFDM subcarriers & $N=16$ \\
Users / targets & $K=5$, $L=4$ \\
Target angles & $[-30^\circ,0^\circ,30^\circ,60^\circ]$ \\
Power budget & $P_{\max}=1$ \\
Noise variance & $\sigma^2=10^{-2}$ \\
Constellation kurtosis & $\kappa=1.32$ \\
User-rate floor & $\gamma_{k,\min}=1$ bps/Hz, $\forall k\in\mathcal{K}$ \\
Illumination floor & $P_{\mathrm{des}}=0.8P_{\max}/N$ \\
CV sweep & $\mathrm{CV}_{\max}\in\{0,0.1,\ldots,0.9\}$ \\
Monte Carlo realizations & $100$ independent channel realizations \\
GR trials & $R_{\mathrm G}=10$ \\
\hline
\end{tabular}
\end{table}
\begin{figure*}[!t]
    \centering
    \includegraphics[width=\textwidth]{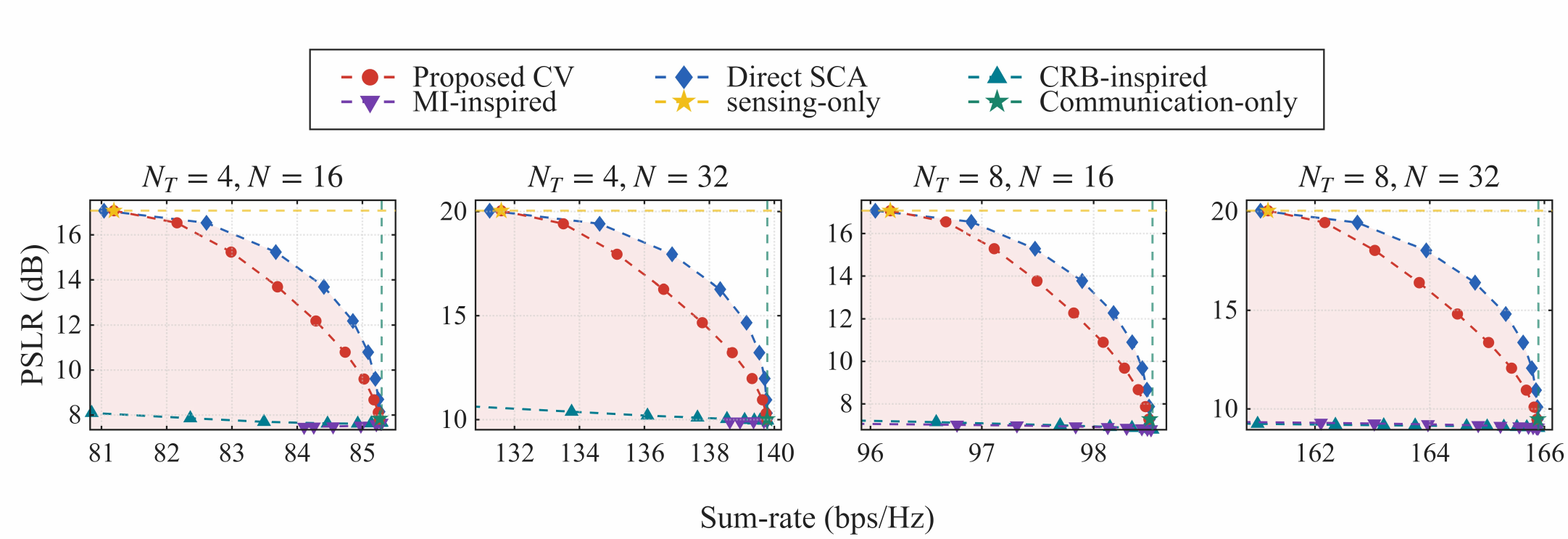}
    \vspace{-1.5em}
    \caption{Sum-rate versus worst-case PSLR Pareto frontiers.}
    \label{fig:pareto_grid}
\end{figure*}
All simulations were run on a 32-core AMD Ryzen Threadripper 9970X workstation using CVX 2.2 and MOSEK 11.2.2.

We compare the proposed CV-constrained AO method against several reference schemes. The communication-only scheme removes the CV and illumination constraints and serves as an upper reference for the sum-rate. The sensing-optimal point corresponds to $\mathrm{CV}_{\max}=0$, which enforces a flat directional power spectrum. The direct PSLR-constrained baseline uses the procedure in Sec.~\ref{subsec:direct_pslr_baseline}. We further include CRB- and MI-inspired beamforming baselines, representative of conventional estimation- and information-theoretic sensing objectives in ISAC \cite{KeskinTSP2021,LiTVT2024_MI}. Let $\mathbf{R}_x=\frac{1}{N}\sum_{n\in\mathcal{N}}\mathbf{W}_n$ denote the average transmit covariance, $\mathbf{B}=[\mathbf{a}^*(\theta_0),\ldots,\mathbf{a}^*(\theta_{L-1})]$, and $\dot{\mathbf{b}}_l=\partial\mathbf{a}^*(\theta_l)/\partial\theta_l$. The CRB-inspired baseline constrains a weighted spatial-Fisher-information utility $\sum_{l\in\mathcal{L}} \dot{\mathbf{b}}_l^{\mathsf{H}}\mathbf{R}_x\dot{\mathbf{b}}_l/\sigma^2$, whereas the MI-inspired baseline constrains the target-subspace log-det utility $\log_2\det(\mathbf{I}+\mathbf{B}^{\mathsf{H}}\mathbf{R}_x\mathbf{B}/\sigma^2)$. The CRB- and MI-inspired baselines use the same settings except that the CV/PSLR constraints are omitted.
\begin{figure}[!t]
    \centering
    \includegraphics[width=\columnwidth]{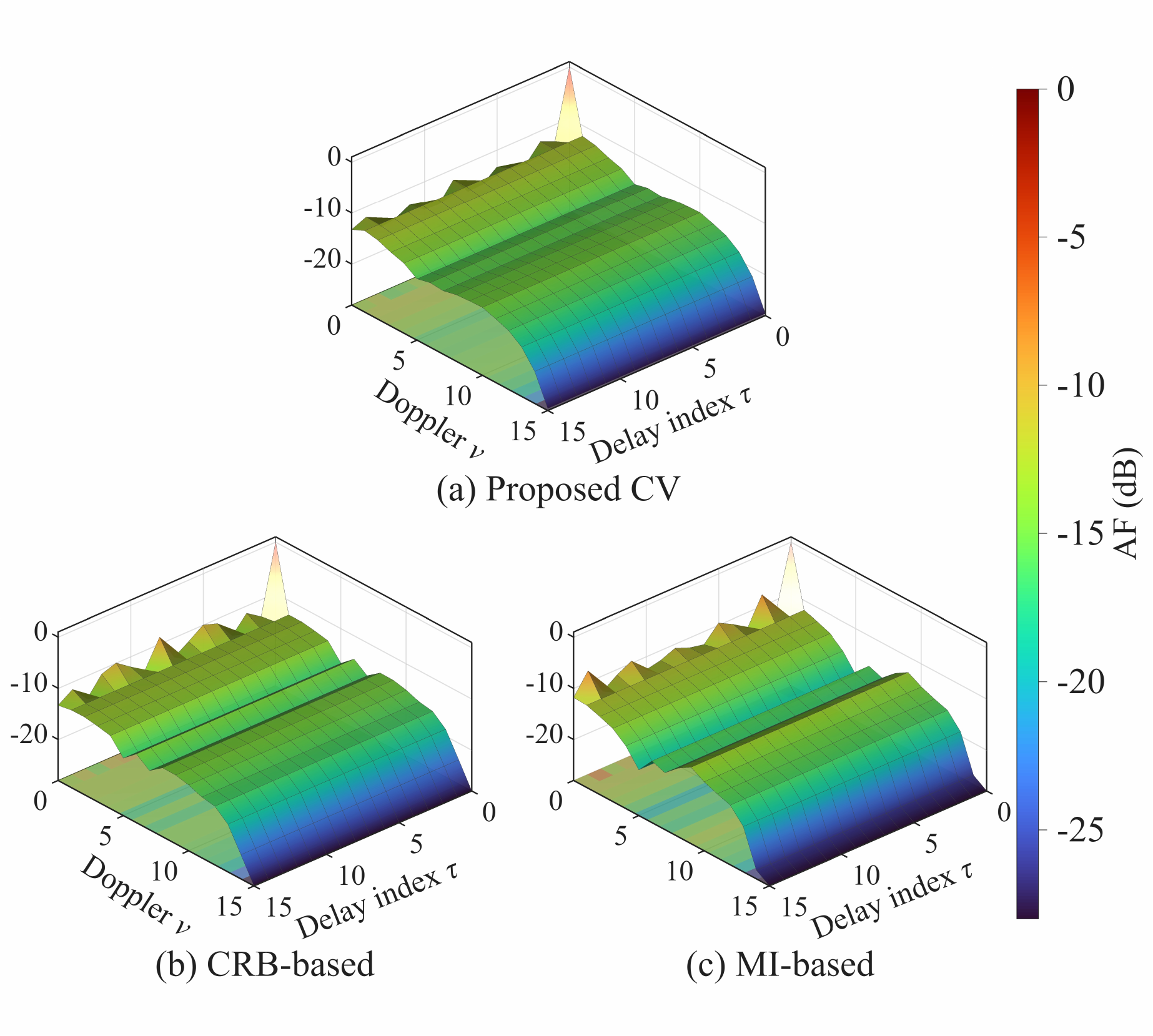}
    \vspace{-1.5em}
    \caption{Normalized delay--Doppler AF surfaces for representative designs.}
    \label{fig:af_3d_comparison}
    \vspace{-1.5em}
\end{figure}

\begin{figure}[!t]
    \centering
    \includegraphics[width=\columnwidth]{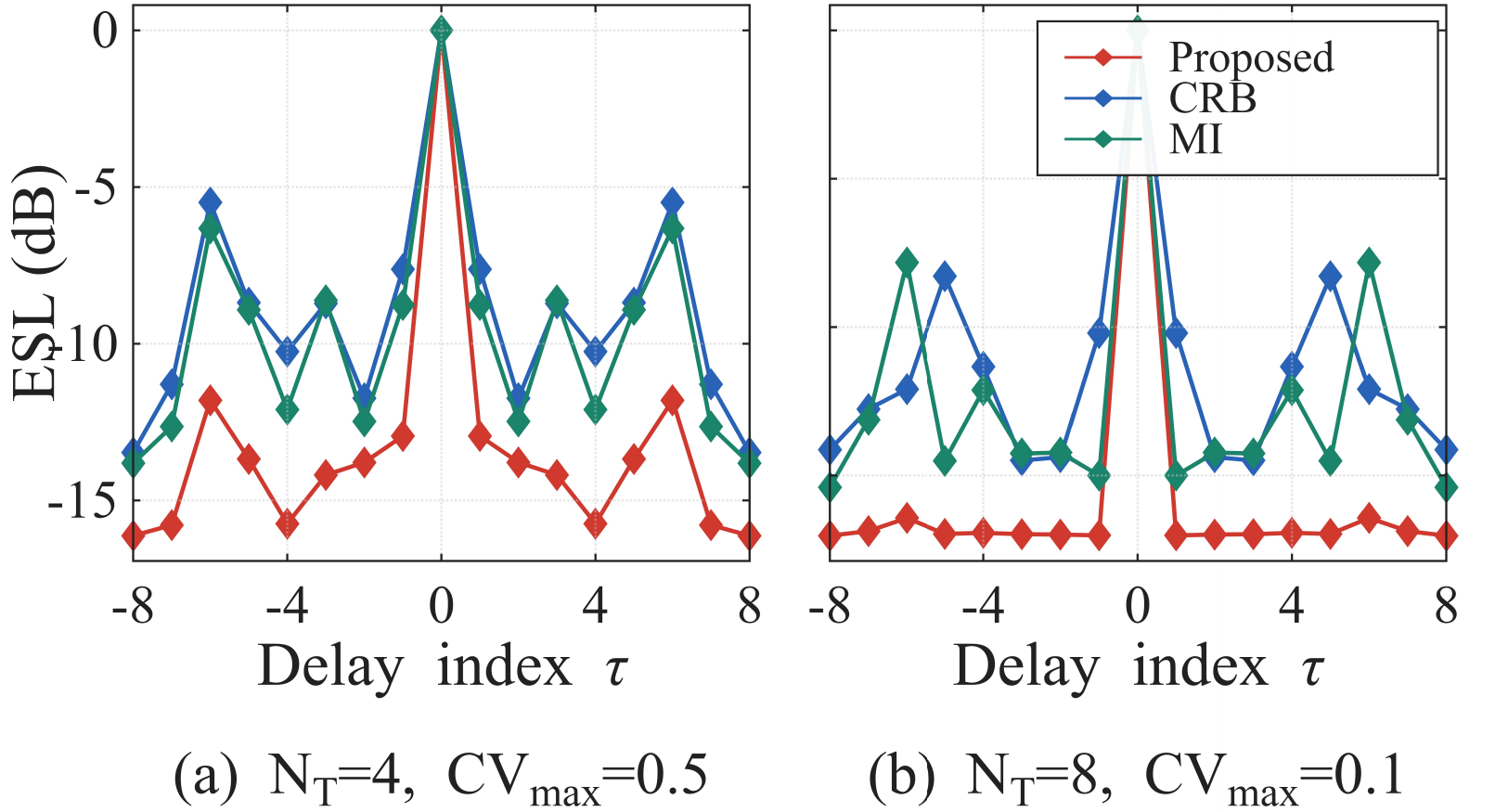}
    \vspace{-1.5em}
    \caption{Worst-direction zero-Doppler ESL cuts for the compared designs.}
    \label{fig:af_worst_cut}
\end{figure}
\subsection{PSLR Pareto Frontiers}
\label{subsec:pslr_pareto}

Fig.~\ref{fig:pareto_grid} shows the sum-rate--PSLR Pareto frontiers averaged over $100$ channel realizations. The four panels correspond to different configurations, and the horizontal dashed line marks the global PSLR limit in Theorem~\ref{thm: PSLR global bound}. Rate differences and percentages are computed within each fixed-$(N_T,N)$ configuration. The key trend is that $\mathrm{CV}_{\max}$ directly controls the sensing--communication tradeoff: at the strict flat-spectrum point, the proposed design raises the worst-case PSLR from the communication-only level of $6.79$--$9.96$ dB to $17.08$ dB for $N=16$ and $20.04$ dB for $N=32$, i.e., by about $9.4$--$11.0$ dB. This PSLR gain costs $2.35$--$8.20$ bps/Hz relative to the communication-only solution, corresponding to approximately $2.4$--$5.9\%$ of the communication-only sum-rate. Conversely, at the loose $\mathrm{CV}_{\max}=0.9$, the rate loss is at most $0.04$ bps/Hz in all four configurations, and the PSLR returns close to the communication-only value. Thus, the proposed CV-constrained design achieves substantial ESL-based zero-Doppler PSLR gains with a limited sum-rate penalty at strict operating points.

The direct PSLR-constrained baseline achieves slightly higher sum-rates than the proposed CV-reformulated scheme. Across four cases, the average sum-rate difference is $0.18$--$0.69$ bps/Hz, and the maximum pointwise average difference is $1.74$ bps/Hz. The proposed curve achieves $99.7\%$ of the direct PSLR-constrained baseline's rate on average. For the conventional baselines, the MI-inspired design remains close to the communication-only rate and reaches a maximum PSLR of $7.36$--$9.96$ dB. While the CRB-inspired design achieves optimal PSLR, its sum-rate drops to $60.40$--$92.90$ bps/Hz. The results show that these baselines yield only marginal PSLR gains. In contrast, the CV directly constrains the subcarrier-wise directional-power flatness that determines the feasible domain of the PSLR.

Figs.~\ref{fig:af_3d_comparison} and~\ref{fig:af_worst_cut} provide representative AF visualizations of this Pareto behavior. Fig.~\ref{fig:af_3d_comparison} compares the proposed, CRB- and MI-inspired designs at comparable sum-rate operating points. The proposed design suppresses localized peaks in the delay--Doppler surface and lowers the range sidelobe envelope in the worst target direction, while the conventional baselines retain pronounced sidelobe peaks. Fig.~\ref{fig:af_worst_cut} compares the range sidelobes of the three designs in the zero-Doppler cut across two realizations.

\subsection{Computational Burden and Feasibility Under System Stress}

\begin{figure}[!t]
    \centering
    \includegraphics[width=\columnwidth]{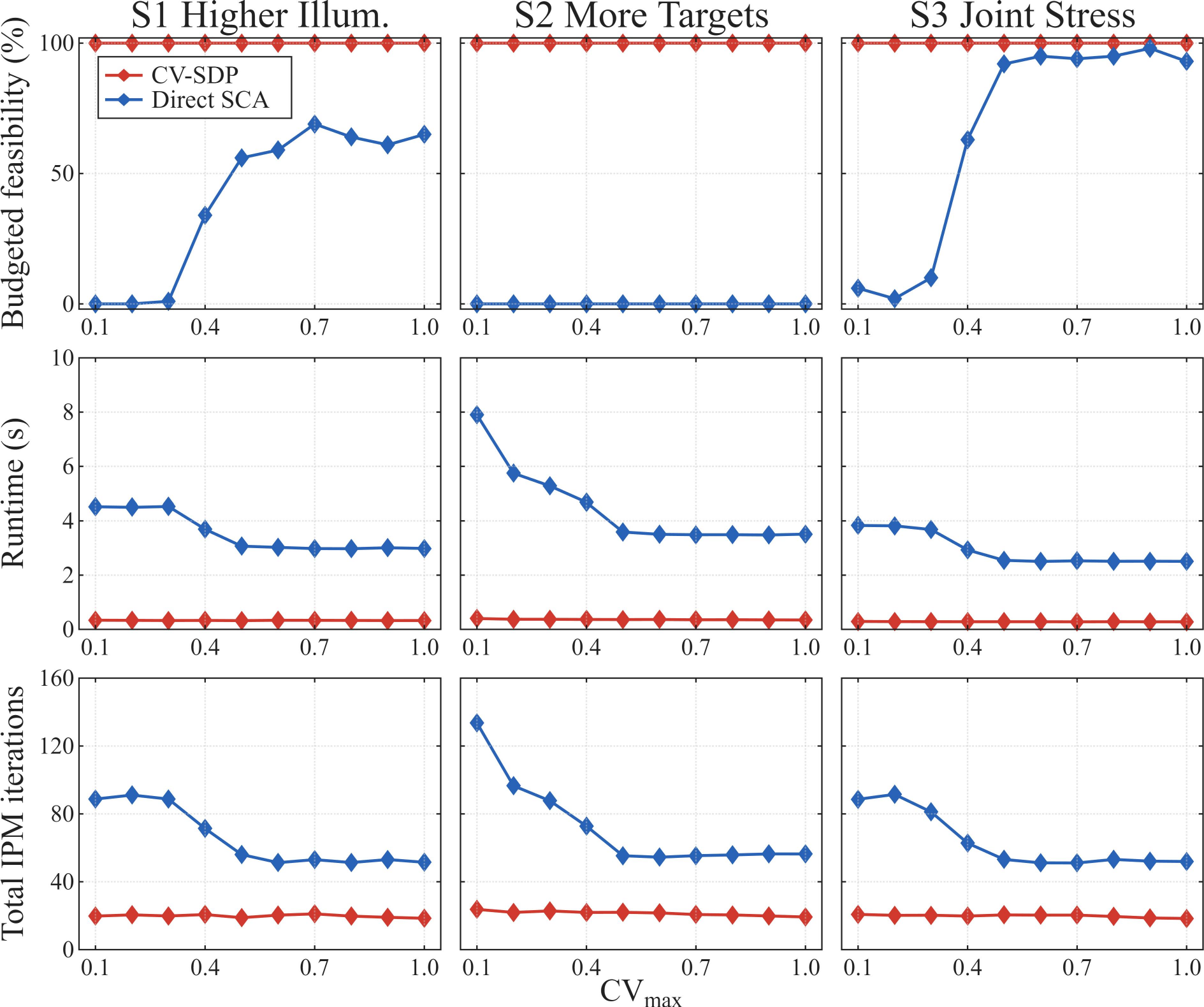}
    \vspace{-1.5em}
    \caption{Feasibility and computational burden under three scenarios.}
    \label{fig:stress_burden}
\end{figure}

Fig.~\ref{fig:stress_burden} compares the proposed CV-based formulation and the direct PSLR-constrained baseline under three stress scenarios. S1 represents a higher illumination requirement with $P_{\mathrm{des}}=1.10P_{\max}/N$. S2 increases the number of target directions to $L=6$. S3 jointly tightens the QoS and illumination requirements using $L=4$, $\gamma_{k,\min}=2.00$, and $P_{\mathrm{des}}=1.10P_{\max}/N$. All scenarios use $N_T=4$, $N=16$, and $\mathrm{CV}_{\max}\in\{0.1,\ldots,1.0\}$. Each point averages $100$ independent channel realizations. The figure reports time-budgeted feasibility together with the wall-clock runtime and accumulated interior-point iterations. Time-budgeted feasibility denotes a feasible solution obtained within 3~s. The 3-s deadline is used as a common computational benchmark rather than as a protocol-level real-time requirement. We set $\mathrm{PSLR}_{\min}=(1-10^{-5})\mathcal L(N,\kappa,\mathrm{CV}_{\max})$.

The proposed CV-based formulation meets the 3-s runtime budget for every tested instance, whereas the direct PSLR-constrained baseline achieves average feasibility rates of only \(40.9\%\), \(0.0\%\), and \(64.8\%\) in S1--S3 and deteriorates as the CV requirement tightens. Across the displayed sweep, the CV-based formulation reduces total runtime by factors of \(10.68\), \(12.15\), and \(10.35\) in S1--S3, respectively. Below \(\mathrm{CV}_{\max}=0.4\), both the runtime and accumulated interior-point method (IPM) iterations of the direct baseline increase sharply. The average wall-clock time per accumulated IPM iteration is also lower for the CV-based formulation: \(16.67\), \(17.03\), and \(14.27\)~ms, compared with \(54.57\), \(62.19\), and \(46.71\)~ms for the direct baseline, making each direct-baseline iteration \(3.27\times\), \(3.65\times\), and \(3.27\times\) more costly. This gain stems from imposing one compact SOC constraint per target and avoiding the nested SCA loop, which repeatedly canonicalizes a denser, delay-coupled conic system; consequently, the CV-based formulation provides higher deadline-constrained feasibility and more predictable computation under stringent sensing requirements.

\subsection{Neural Network-Based Evaluation of Constraint Representations}
\begin{figure}[!t]
    \centering
    \includegraphics[width=\columnwidth]{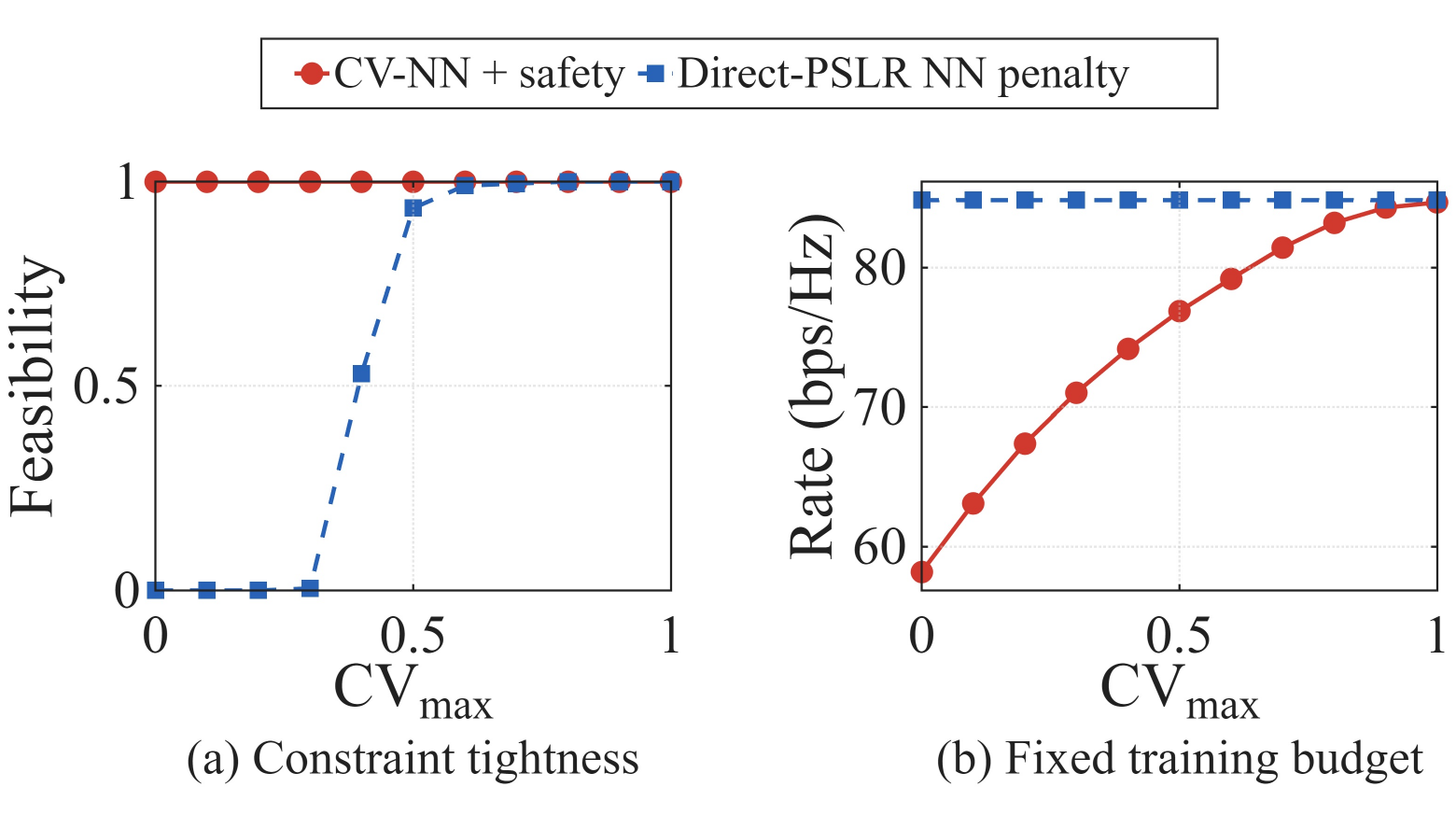}
    \vspace{-1.5em}
    \caption{NN-based constraint enforcement: (a) feasibility of fixed covariance outputs under a threshold sweep and (b) final-epoch sum-rate under a fixed training budget.}
    \label{fig:ml_cv_advantage}
\end{figure}
We next examine how CV and direct PSLR constraints can be incorporated into a learning-based beamforming model. To isolate the constraint-enforcement mechanism, two label-free covariance-factor neural networks (NNs) share the same backbone, training budget, and positive-semidefinite (PSD) output structure.

The CV constraint enables the simple closed-form feasibility layer derived below, whereas the direct PSLR constraint admits no analogous closed-form correction and is therefore handled through a training penalty. Following constrained-learning architectures \cite{AgrawalNeurIPS2019,TordesillasRAYEN2023,LiTWC2021,RashwanNeurIPS2025}, we implement the CV correction as a deterministic output layer.

Let \(\widehat{\mathbf W}_n\succeq\mathbf 0\) denote the raw covariance predicted on subcarrier \(n\), and define its subcarrier average as
\begin{equation}
    \overline{\mathbf W}=\frac{1}{N}\sum_{n\in\mathcal{N}}\widehat{\mathbf W}_n.
    \label{eq:nn_cv_flat_anchor}
\end{equation}
For target \(l\), let \(\widehat p_{l,n}=\operatorname{Tr}(\mathbf A_l\widehat{\mathbf W}_n)\), with mean \(\widehat\mu_l\) and standard deviation \(\widehat\sigma_l\) over the subcarriers. Repeating \(\overline{\mathbf W}\) on every subcarrier gives a flat directional-power profile with the same mean. The corrected covariance is
\begin{align}
    \mathbf W_n^{\mathrm{safe}}
    &=\rho\widehat{\mathbf W}_n+(1-\rho)\overline{\mathbf W},
    \label{eq:nn_cv_safety_layer}\\
    \rho
    &=\min\!\left\{1,\min_l \frac{\mathrm{CV}_{\max}\widehat\mu_l}{\widehat\sigma_l}\right\},
    \label{eq:nn_cv_safety_rho}
\end{align}
where the ratio is taken as unbounded when \(\widehat\sigma_l=0\), so a flat profile imposes no restriction on \(\rho\). Since \(\operatorname{Tr}(\mathbf A_l\overline{\mathbf W})=\widehat\mu_l\), the corrected directional-power profile satisfies
\begin{equation}
    p_{l,n}^{\mathrm{safe}}
    =\widehat\mu_l+\rho(\widehat p_{l,n}-\widehat\mu_l),
    \quad
    \mathrm{CV}(\mathbf p_l^{\mathrm{safe}})
    =\rho\frac{\widehat\sigma_l}{\widehat\mu_l}
    \leq\mathrm{CV}_{\max}.
    \label{eq:nn_cv_safety_guarantee}
\end{equation}
As a convex combination of PSD matrices, \eqref{eq:nn_cv_safety_layer} guarantees positive semidefiniteness while strictly preserving both total transmit power and target illumination levels:
\begin{equation}
    \frac{1}{N}\sum_{n\in\mathcal{N}}p_{l,n}^{\mathrm{safe}}=\widehat\mu_l.
    \label{eq:nn_cv_illum_preserve}
\end{equation}
Thus, the layer enforces the CV constraint without altering the NN's predicted illumination level.

Fig.~\ref{fig:ml_cv_advantage} considers $N_T=4$ and $N=16$, with $8{,}800$ training and $2{,}200$ validation channel realizations. By varying the requested threshold for fixed covariance predictions, Fig.~\ref{fig:ml_cv_advantage}(a) evaluates constraint enforcement independently of network scale or training time. Fig.~\ref{fig:ml_cv_advantage}(b) reports the final-epoch sum-rate under a fixed training budget.

The CV model satisfies the constraint across all $2{,}200$ validation channels, confirming \eqref{eq:nn_cv_safety_guarantee}. In contrast, the direct PSLR model yields near-zero feasibility for $\mathrm{CV}_{\max}\leq0.3$, which gradually increases to $100\%$ for $\mathrm{CV}_{\max}\geq0.8$. Consequently, its nominal sum-rate of $84.85$~bps/Hz at low $\mathrm{CV}_{\max}$ is practically unachievable. Conversely, the CV model provides feasible sum-rates scaling from $58.19$ to $84.68$~bps/Hz. This $100\%$ feasibility stems from the deterministic safety layer rather than learned generalization, highlighting the CV representation as an effective explicit constraint interface for both optimization and learning frameworks.

\subsection{Impact of the Illumination Constraint}
\label{subsec:illumination_constraint}

We next investigate the effect of the illumination floor $P_{\mathrm{des}}$ in \eqref{const:p2f_illum}. To isolate this effect, we use a two-target setup with $\theta=[0^\circ,30^\circ]$, $N_T=8$, $N=32$, and $\gamma_{k,\min}=0$ for all users. The illumination floor is swept as $P_{\mathrm{des}}=nP_{\max}/N$ with $n\in\{0,1,2,3\}$.

\begin{figure}[t]
    \centering
    \includegraphics[width=\columnwidth]{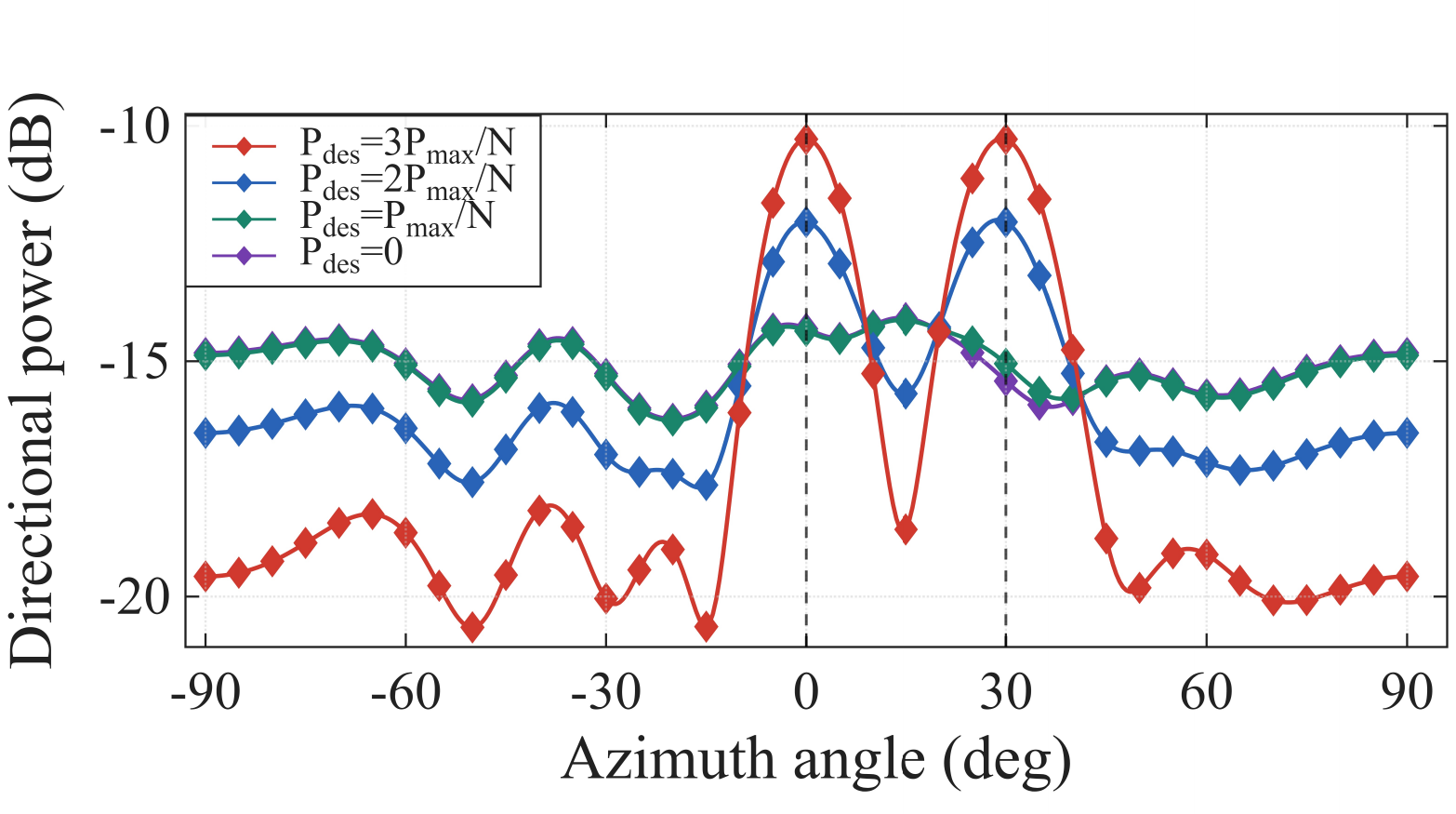}
    \vspace{-1.5em}
    \caption{Directional-power patterns under different illumination floors.}
    \label{fig:beamgain_pdes_sweep}
    \vspace{-1.5em}
\end{figure}

Fig.~\ref{fig:beamgain_pdes_sweep} shows one representative channel realization at $\mathrm{CV}_{\max}=0.5$, with vertical dashed lines marking the target directions. Increasing $P_{\mathrm{des}}$ forces the transmit covariance to allocate more directional power toward these angles. For $n=0$ and $n=1$, the optimized beampatterns are nearly identical because the mild illumination requirement is already satisfied by the rate-oriented solution. As $n$ increases to $2$ and $3$, clear beams are formed at the target directions, while the off-target directional power is reduced due to the fixed total transmit-power budget.

\begin{figure}[t]
    \centering
    \includegraphics[width=\columnwidth]{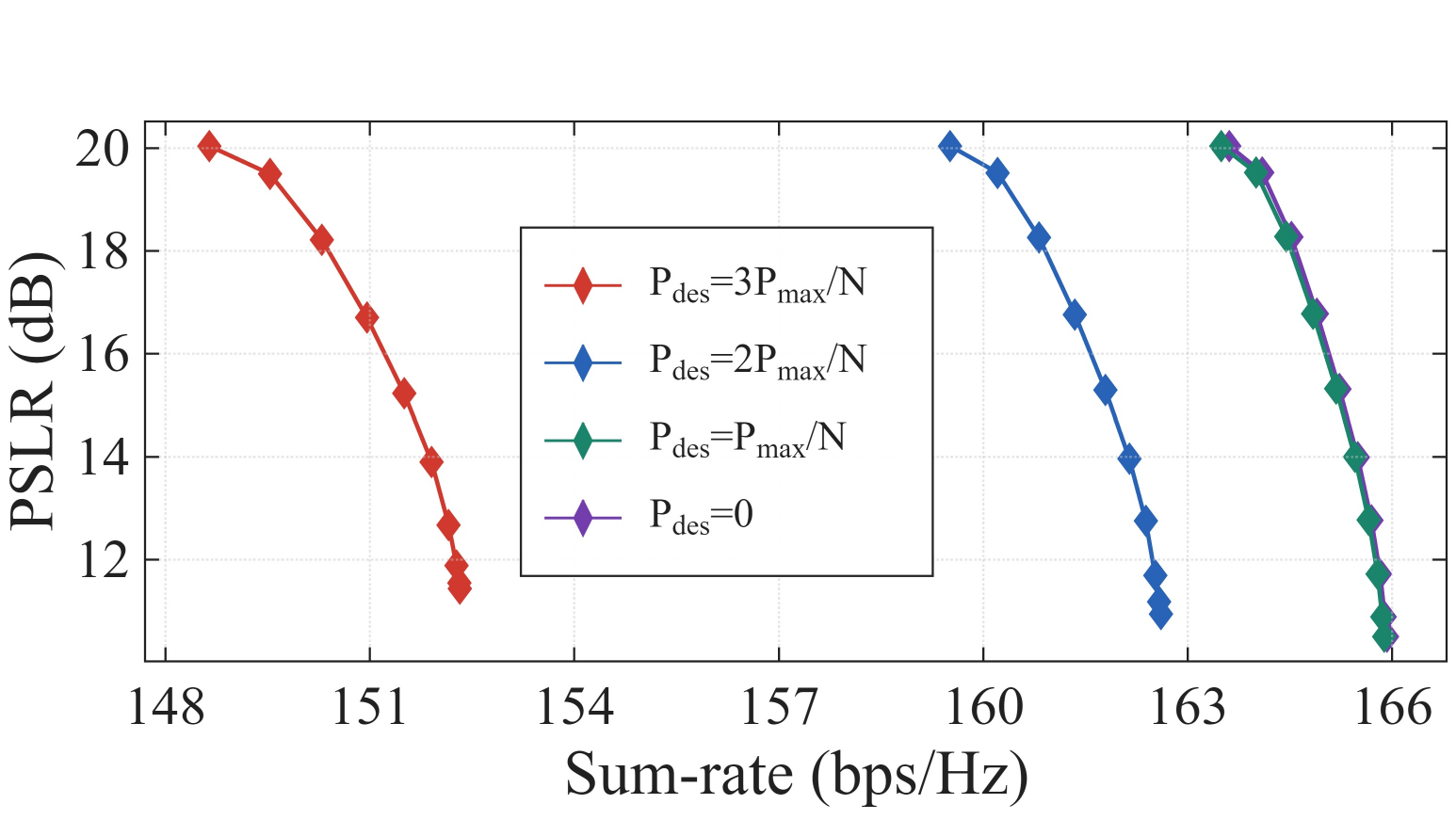}
    \vspace{-1.5em}
    \caption{Sum-rate--PSLR tradeoffs under different illumination floors.}
    \label{fig:pdes_pareto_sweep}
    \vspace{-1.5em}
\end{figure}

Fig.~\ref{fig:pdes_pareto_sweep} averages each curve over ten channel realizations and sweeps $\mathrm{CV}_{\max}\in\{0,0.1,\ldots,0.9\}$. A tighter illumination floor shifts the entire Pareto curve leftward. The sum-rate range changes from $163.61$--$165.93$ bps/Hz for $n=0$ and $163.50$--$165.89$ bps/Hz for $n=1$, to $159.51$--$162.60$ bps/Hz for $n=2$ and $148.64$--$152.32$ bps/Hz for $n=3$. This confirms that $P_{\mathrm{des}}$ controls the transmit-side illumination floor. Tightening this proxy increases directional transmit power but consumes power and spatial DoF otherwise available to communication. The CV and illumination constraints therefore play complementary roles: CV controls the subcarrier-wise power variation and hence the AF shape, whereas $P_{\mathrm{des}}$ controls the mean directional energy toward the targets.


\section{Conclusion}

This paper established an AF-aware design framework for MUMT MIMO--OFDM ISAC. From the derived DP-AF and closed-form ESL, we showed that the absence of delay--Doppler coupling and full two-dimensional ISLR invariance persist under MIMO beamforming, whereas the zero-Doppler PSLR remains controllable. This result motivated a joint beamforming and subcarrier-allocation formulation incorporating PSLR constraint. We further proved that, for symbol kurtosis greater than one, the global PSLR maximum is achieved if and only if each target-wise directional power profile is flat across the subcarriers. Tight CV-dependent PSLR bounds then converted this condition into a single convex sufficient constraint and an efficient alternating-optimization method.

At the flat-spectrum point, the proposed design improved the PSLR by $9.4$--$11.0$ dB over communication-only beamforming with only a $2.4$--$5.9\%$ sum-rate loss. It also attained $99.7\%$ of the direct PSLR-constrained baseline's average sum-rate while reducing runtime by factors of $10.35$--$12.15$. The NN evaluation further demonstrated that the convex CV constraint enables a deterministic safety layer, thereby guaranteeing sensing feasibility by construction. These results establish the CV as a theoretically grounded and computationally efficient interface for AF-aware MUMT ISAC design. Future work will extend the framework to joint range--Doppler--angle sidelobe control and dynamic ISAC scenarios.

\begin{appendices}
\section{Proof of Lemma~\ref{lem:ESL}}
\label{app:proof_ESL}

To derive the expected sidelobe level $\mathrm{ESL}[\tau,\nu;\theta] = \mathbb{E}[|\chi[\tau,\nu;\theta]|^2]$, we first express the discrete MIMO ambiguity function in a mathematically convenient form. Let $\tilde{\mathbf{s}} = \mathrm{diag}(\mathbf{W}\mathbf{a}(\theta))\mathbf{s}$. The AF can be rewritten as
\begin{equation}
    \chi[\tau, \nu; \theta] = \tilde{\mathbf{s}}^{\mathsf{H}} \mathbf{F}_N \mathbf{J}^{\mathsf{T}}_{N,\tau} \mathbf{D}^*_{N,\nu} \mathbf{F}_N^{\mathsf{H}} \tilde{\mathbf{s}}.
\end{equation}
By expressing the unitary DFT matrix as $\mathbf{F}_N = [\mathbf{f}_0, \mathbf{f}_1, \dots, \mathbf{f}_{N-1}]$, the AF expands to
\begin{subequations}
\begin{align}
    \chi[\tau, \nu; \theta] &= \tilde{\mathbf{s}}^{\mathsf{H}} \begin{bmatrix} \mathbf{f}_0 & \cdots & \mathbf{f}_{N-1} \end{bmatrix} \mathbf{J}^{\mathsf{T}}_{N,\tau} \mathbf{D}^*_{N,\nu} \begin{bmatrix} \mathbf{f}^{\mathsf{H}}_0 \\ \vdots \\ \mathbf{f}^{\mathsf{H}}_{N-1} \end{bmatrix} \tilde{\mathbf{s}} \\
    &= \tilde{\mathbf{s}}^{\mathsf{H}} \begin{bmatrix} \mathbf{f}_0 & \cdots & \mathbf{f}_{N-1} \end{bmatrix} \begin{bmatrix} \mathbf{f}^{\mathsf{H}}_\tau \cdot e^{-j2\pi\frac{\tau\nu}{N}} \\ \vdots \\ \mathbf{f}^{\mathsf{H}}_{\tau-1} \cdot e^{-j2\pi\frac{(\tau-1)\nu}{N}} \end{bmatrix} \tilde{\mathbf{s}} \\
    &= \sum_{n\in\mathcal{N}} e^{-j2\pi\frac{(n+\tau)\nu}{N}} \tilde{\mathbf{s}}^{\mathsf{H}} \mathbf{f}_n \mathbf{f}^{\mathsf{H}}_{n+\tau} \tilde{\mathbf{s}}.
\end{align}
\end{subequations}

Therefore, the ESL can be formulated by taking the expectation of the magnitude squared:
\begin{subequations}
\begin{align}
    \mathrm{ESL} &= \mathbb{E}\left[ \left| \sum_{n\in\mathcal{N}} e^{-j2\pi\frac{n\nu}{N}} \tilde{\mathbf{s}}^{\mathsf{H}} \mathbf{f}_{n-\tau} \mathbf{f}^{\mathsf{H}}_n \tilde{\mathbf{s}} \right|^2 \right] \\
    &= \mathbb{E}\left[ \sum_{n\in\mathcal{N}}\sum_{m\in\mathcal{N}} e^{-j2\pi\frac{(n-m)\nu}{N}} \mathbf{f}^{\mathsf{H}}_{m-\tau} \tilde{\mathbf{s}} \tilde{\mathbf{s}}^{\mathsf{H}} \mathbf{f}_m \mathbf{f}^{\mathsf{H}}_n \tilde{\mathbf{s}} \tilde{\mathbf{s}}^{\mathsf{H}} \mathbf{f}_{n-\tau} \right].
\end{align}
\end{subequations}

To handle the expectation over the random data symbols, we utilize the Kronecker product. Let $\mathbf{s}_0 = \tilde{\mathbf{s}}^* \otimes \tilde{\mathbf{s}} \in \mathbb{C}^{N^2}$. Using the property $\mathbf{x}^{\mathsf{H}} \mathbf{A} \mathbf{y} = (\mathbf{y}^{\mathsf{T}} \otimes \mathbf{x}^{\mathsf{H}})\mathrm{vec}(\mathbf{A})$, we have $\mathbf{f}^{\mathsf{H}}_{m-\tau} \tilde{\mathbf{s}} \tilde{\mathbf{s}}^{\mathsf{H}} \mathbf{f}_m = (\mathbf{f}^{\mathsf{T}}_m \otimes \mathbf{f}^{\mathsf{H}}_{m-\tau})\mathbf{s}_0$, which leads to
\begin{equation} \label{eq:app_ESL_kron}
    \mathrm{ESL} = \sum_{n\in\mathcal{N}}\sum_{m\in\mathcal{N}} e^{-j2\pi\frac{(n-m)\nu}{N}} (\mathbf{f}^{\mathsf{T}}_m \otimes \mathbf{f}^{\mathsf{H}}_{m-\tau}) \mathbb{E}[\mathbf{s}_0 \mathbf{s}^{\mathsf{H}}_0] (\mathbf{f}^*_n \otimes \mathbf{f}_{n-\tau}).
\end{equation}

Next, we evaluate the covariance matrix $\mathbb{E}[\mathbf{s}_0 \mathbf{s}^{\mathsf{H}}_0]$. Since $\mathbf{s}_0 = [w_0^*(\theta) s_0^* \tilde{\mathbf{s}}^{\mathsf{T}}, \dots, w_{N-1}^*(\theta) s_{N-1}^* \tilde{\mathbf{s}}^{\mathsf{T}}]^{\mathsf{T}}$, the expected value of the $(i, j)$-th element in the $(a, b)$-th block of $\mathbf{s}_0 \mathbf{s}^{\mathsf{H}}_0$ is given by
\begin{equation}
    w_a^*(\theta) w_b(\theta) w_i(\theta) w_j^*(\theta) \mathbb{E}[s_a^* s_b s_i s_j^*].
\end{equation}
Because the data symbols $\{s_k\}$ are mutually independent with zero mean, unit variance, and kurtosis $\kappa$, the expectation $\mathbb{E}[s_a^* s_b s_i s_j^*]$ is non-zero only under specific index conditions:
\begin{itemize}
    \item \textbf{Case I ($a \neq b$):} The expectation is non-zero if and only if $a=i$ and $b=j$. The corresponding term reduces to $P_a(\theta)P_b(\theta)$.
    \item \textbf{Case II ($a = b$):} The expectation is non-zero if and only if $i=j$. If $a=i$, it yields $\kappa P_a^2(\theta)$. If $a \neq i$, it yields $P_a(\theta)P_i(\theta)$.
\end{itemize}

Substituting these non-zero elements back into \eqref{eq:app_ESL_kron}, the element corresponding to Case I is multiplied by $[\mathbf{f}_m]_a [\mathbf{f}_{m-\tau}]_a^* [\mathbf{f}_n]_b^* [\mathbf{f}_{n-\tau}]_b = \frac{1}{N^2}e^{-j2\pi\frac{(a-b)\tau}{N}}$. For Case II, it is multiplied by $\frac{1}{N^2}e^{-j2\pi\frac{(m-n)(a-i)}{N}}$.

Accumulating all non-zero terms, the ESL expands to
\begin{equation} \label{eq:app_ESL_expand}
\begin{aligned}
    &\mathrm{ESL}
    = N^{-2}\!\sum_{n,m\in\mathcal{N}}\! e^{-j2\pi\frac{(n-m)\nu}{N}} \\
    &~ \times \Bigg[\! \sum_{a\in\mathcal{N}}\!\sum_{b\in\mathcal{N}}\! P_a(\theta)P_b(\theta) e^{-j2\pi\frac{(a-b)\tau}{N}}
    + (\kappa \!-\! 2)\!\sum_{a\in\mathcal{N}}\! P_a^2(\theta) \\
    &~~ + \sum_{a\in\mathcal{N}}\!\sum_{i\in\mathcal{N}}\! P_a(\theta)P_i(\theta) e^{-j2\pi\frac{(m-n)(a-i)}{N}} \!\Bigg]\!.
\end{aligned}
\end{equation}

We can evaluate the double summation over $m$ and $n$ for the first two terms in the bracket using the identity $\sum_{n\in\mathcal{N}}\sum_{m\in\mathcal{N}} e^{-j2\pi\frac{(n-m)\nu}{N}} = N^2 \delta_{\nu,0}$.
For the third term in the bracket, reorganizing the exponentials yields:
\begin{subequations}
\begin{align}
    &\sum_{n,m,a,i} P_a(\theta)P_i(\theta) e^{-j2\pi\frac{(m-n)(a-i-\nu)}{N}} \\
    &\hspace{-1mm}=\!\!\!\sum_{a\in\mathcal{N}}\!\!\sum_{i\in\mathcal{N}}\!\!P_a(\theta)P_i(\theta)\!\!\left( \sum_{m\in\mathcal{N}}\!e^{-j2\pi\frac{m(a-i-\nu)}{N}}\!\!\right) \!\!\!\left( \sum_{n\in\mathcal{N}}\!e^{j2\pi\frac{n(a-i-\nu)}{N}}\!\!\right)\!\!.
\end{align}
\end{subequations}
The summations over $m$ and $n$ are non-zero (equal to $N$) if and only if $a-i-\nu \equiv 0 \pmod{N}$. Since $0\leq a,i,\nu\leq N-1$, this condition admits the two cases $a-i-\nu=0$ and $a-i-\nu=-N$. The former gives $i=a-\nu$ for $a=\nu,\ldots,N-1$, whereas the latter gives $i=a+N-\nu$ for $a=0,\ldots,\nu-1$. Thus, the double summation in the third term evaluates to
\begin{equation}
    N^2\!\left[\sum_{a=\nu}^{N-1} P_a(\theta)P_{a-\nu}(\theta)
    +\sum_{a=0}^{\nu-1} P_a(\theta)P_{a+N-\nu}(\theta)\right].
\end{equation}
After the $1/N^2$ prefactor in \eqref{eq:app_ESL_expand} is applied, its contribution is the circular-correlation form $\sum_{a\in\mathcal{N}} P_a(\theta)P_{a-\nu}(\theta)$.

Applying these simplifications to \eqref{eq:app_ESL_expand} gives the final closed-form expression:
\begin{equation}
\begin{aligned}
    \mathrm{ESL}
    &= \delta_{\nu,0} \Bigg[ \sum_{a\in\mathcal{N}}\sum_{b\in\mathcal{N}} P_a(\theta)P_b(\theta) e^{-j2\pi\frac{(a-b)\tau}{N}} \\
    &~ + (\kappa - 2)\sum_{a\in\mathcal{N}} P_a^2(\theta) \Bigg] + \sum_{a\in\mathcal{N}} P_a(\theta)P_{a-\nu}(\theta),
\end{aligned}
\end{equation}
which concludes the proof. \hfill $\blacksquare$

\section{Derivation of the Total AF Energy}
\label{app:deriv_ISL}

Let $Q(\theta)\triangleq\sum_{n\in\mathcal{N}}P_n(\theta)^2$ and $S(\theta)\triangleq\sum_{n\in\mathcal{N}}P_n(\theta)$. Using the closed-form ESL in Lemma~\ref{lem:ESL}, the total expected AF energy is
\begin{align}
E_{\mathrm{total}}
&\triangleq \sum_{\tau\in\mathcal{N}}\sum_{\nu\in\mathcal{N}}\mathrm{ESL}[\tau,\nu;\theta] \nonumber\\
&= \sum_{\tau\in\mathcal{N}}\left[\left|\sum_{n\in\mathcal{N}}P_n(\theta)e^{j2\pi n\tau/N}\right|^2+(\kappa-2)Q(\theta)\right] \nonumber\\
&~ +N\sum_{\nu\in\mathcal{N}}\sum_{a\in\mathcal{N}}P_a(\theta)P_{a-\nu}(\theta).
\end{align}
By DFT orthogonality and circular indexing,
\begin{align}
\sum_{\tau\in\mathcal{N}}\left|\sum_{n\in\mathcal{N}}P_n(\theta)e^{j2\pi n\tau/N}\right|^2
&=NQ(\theta), \nonumber\\
\sum_{\nu\in\mathcal{N}}\sum_{a\in\mathcal{N}}P_a(\theta)P_{a-\nu}(\theta)
&=S(\theta)^2.
\end{align}
Consequently,
\begin{align}
E_{\mathrm{total}}
&=NQ(\theta)+N(\kappa-2)Q(\theta)+NS(\theta)^2 \nonumber\\
&=N\big[(\kappa-1)Q(\theta)+S(\theta)^2\big] \nonumber\\
&=N\,\mathrm{ESL}[0,0;\theta],
\end{align}
which proves \eqref{eq:AF_total_energy}. Thus, the full two-dimensional ISLR equals $N-1$ for every directional power profile and cannot be reduced by transmit beamforming. \hfill $\blacksquare$

\section{Proof of Theorem~\ref{thm: PSLR global bound}}
\label{app:proof_thm1}
Using the mainlobe and peak-sidelobe terms defined above, the target-wise PSLR is
\begin{align}
    \mathrm{PSLR}_l
    &=\frac{\mathcal{N}_{\mathrm{P}}(\mathbf{p}_l)}{\mathcal{D}_{\mathrm{P}}(\mathbf{p}_l)} \nonumber\\
    &=\frac{|R_{l,0}|^2+(\kappa-1)\|\mathbf{p}_l\|_2^2}{\max_{\tau\in\mathcal{N}\setminus\{0\}}|R_{l,\tau}|^2+(\kappa-1)\|\mathbf{p}_l\|_2^2}.
\end{align}
Since the peak-sidelobe term is non-negative,
\begin{equation}
    \mathrm{PSLR}_l \leq 1+\frac{|R_{l,0}|^2}{(\kappa-1)\|\mathbf{p}_l\|_2^2}.
\end{equation}
Moreover, the Cauchy--Schwarz inequality gives $|R_{l,0}|^2=(\mathbf{1}^{\mathsf{T}}\mathbf{p}_l)^2\leq N\|\mathbf{p}_l\|_2^2$. Hence,
\begin{equation}
    \mathrm{PSLR}_l \leq 1+\frac{N}{\kappa-1}.
\end{equation}
\textit{Equality Condition:} Equality holds if and only if $p_{l,0}=p_{l,1}=\cdots=p_{l,N-1}$, which also yields $R_{l,\tau}=0$ for every $\tau\neq0$. Thus, the global maximum is achieved exclusively by a flat target-wise directional-power profile ($\mathrm{CV}_l=0$).

\section{Proof of Theorem~\ref{thm:cv_pslr_bound}} \label{app:proof_thm3}

We define the target-wise zero-mean deviation sequence as $\delta_{l,n}=p_{l,n}-\mu_l$. For any integer delay $m\neq0$, the DFT of the constant mean $\mu_l$ is zero. Hence, the delay-domain sidelobes are determined by the $N$-point DFT of $\{\delta_{l,n}\}$:
\begin{equation}
    X_l[m]=\sum_{n\in\mathcal N}p_{l,n}e^{j2\pi nm/N}
    =\sum_{n\in\mathcal N}\delta_{l,n}e^{j2\pi nm/N}.
\end{equation}
By Parseval's theorem, the total sidelobe energy evaluates to
\begin{equation}
    \sum_{m\in\mathcal N\setminus\{0\}}|X_l[m]|^2
    =N\sum_{n\in\mathcal N}\delta_{l,n}^2=N^2\sigma_l^2.
\end{equation}
Substituting $\sum_{n\in\mathcal N}p_{l,n}^2=N(\mu_l^2+\sigma_l^2)$ and $\big(\sum_{n\in\mathcal N}p_{l,n}\big)^2=N^2\mu_l^2$ into the PSLR definition gives
\begin{equation} \label{eq:pslr_variance}
    \mathrm{PSLR}_l
    =\frac{(\kappa-1)N(\mu_l^2+\sigma_l^2)+N^2\mu_l^2}
    {(\kappa-1)N(\mu_l^2+\sigma_l^2)+
    \max_{m\in\mathcal N\setminus\{0\}}|X_l[m]|^2}.
\end{equation}

\textit{1) Upper Bound ($\mathcal{U}$):} To upper-bound the PSLR, we minimize the denominator term $\max_{m\in\mathcal{N}\setminus\{0\}} |X_l[m]|^2$. The global minimum of the maximum peak occurs when the total sidelobe energy is uniformly distributed across all $N-1$ non-zero delay bins, yielding:
\begin{equation}
    \max_{m\in\mathcal N\setminus\{0\}}|X_l[m]|^2
    \geq\frac{N^2\sigma_l^2}{N-1}.
\end{equation}
Substituting this minimum peak into \eqref{eq:pslr_variance} and dividing the numerator and denominator by $N\mu_l^2$ yields $\mathcal U(N,\kappa,\mathrm{CV}_l)$ using \eqref{eq:cv_definition}.

\emph{Equality Condition:} The upper bound is achieved when the deviation is an impulse in the frequency domain, e.g., $p_{l,0}=\mu_l+\sigma_l\sqrt{N-1}$ and $p_{l,n}=\mu_l-\sigma_l/\sqrt{N-1}$ for $n\neq0$. This profile is nonnegative provided $\mathrm{CV}_l\leq\sqrt{N-1}$.

\textit{2) Lower Bound ($\mathcal{L}$):} Conversely, to lower-bound the PSLR, we maximize $\max_{m\in\mathcal{N}\setminus\{0\}} |X_l[m]|^2$. Because $\{\delta_{l,n}\}$ is strictly real, its DFT exhibits conjugate symmetry, i.e., $|X_l[m]|^2 = |X_l[N - m]|^2$. The maximum energy concentration therefore depends on the parity of $N$:
\begin{itemize}
    \item \textbf{Even $N$:} All sidelobe energy can perfectly concentrate at the single self-conjugate Nyquist bin ($m = N/2$), yielding $\max_{m\in\mathcal{N}\setminus\{0\}} |X_l[m]|^2 \leq N^2\sigma_l^2$.
    \item \textbf{Odd $N$:} No non-zero self-conjugate bin exists. The energy must be split equally between at least one conjugate pair, yielding $\max_{m\in\mathcal{N}\setminus\{0\}} |X_l[m]|^2 \leq \frac{N^2}{2}\sigma_l^2$.
\end{itemize}
Let $c = 1$ for even $N$ and $c = 1/2$ for odd $N$. Substituting the maximum possible peak $c N^2\sigma_l^2$ into \eqref{eq:pslr_variance} and reformulating with $\mathrm{CV}_l$ yields:
\begin{equation}
    \mathrm{PSLR}_l \geq \frac{N + \kappa - 1 + (\kappa - 1)\mathrm{CV}_l^2}{\kappa - 1 + (cN + \kappa - 1)\mathrm{CV}_l^2}.
\end{equation}
Since the number of subcarriers $N$ in practical OFDM systems is typically an even integer, we proceed with $c = 1$. Substituting $c = 1$ yields the final lower bound
$\mathcal{L}(N, \kappa, \mathrm{CV}_l)$.

\emph{Equality Condition:} The lower bound is achieved when the variance energy is maximally concentrated in the delay domain. The inverse DFT gives the alternating profile $p_{l,n}=\mu_l+\sigma_l(-1)^n$ for even $N$ (valid for $\mathrm{CV}_l\leq1$) and a sinusoidal profile for odd $N$ (valid for $\mathrm{CV}_l\leq1/\sqrt{2}$). This completes the proof.

\end{appendices}

\bibliographystyle{IEEEtran}
\bibliography{refs}

\end{document}